\documentclass[nonacm,sigconf,bookmarksnumbered,unicode,bookmarks=false,hidelinks]{acmart}

\AtBeginDocument{%
  \providecommand\BibTeX{{%
    \normalfont B\kern-0.5em{\scshape i\kern-0.25em b}\kern-0.8em\TeX}}}

\setcopyright{acmcopyright}
\copyrightyear{2024}
\acmYear{2024}
\acmDOI{XXXXXXX.XXXXXXX}

\acmConference[MSR 2024]{21st International Conference on Mining Software Repositories}{April 2024}{Lisbon, Portugal}

\usepackage[frozencache=true,cachedir=minted-cache]{minted}
\usepackage[normalem]{ulem}
\newif\ifusecolor
\usecolortrue
\ifusecolor

  \def\removed#1{{\color{gray} \sout{#1}}}
\else

  \def\removed#1{}
\fi

\usepackage{multirow}
\usepackage{longtable}
\usepackage{dirtree}
\usepackage[most]{tcolorbox}
\usepackage[labelfont=bf, font=bf]{caption}
\usepackage{subcaption}
\usepackage{booktabs}
\usepackage{array}
\usepackage{longtable}

\usepackage{listings}

\definecolor{cloudwhite}{cmyk}{0,0,0,0.025}
\usepackage[inline]{enumitem}
\usepackage{wrapfig}
\usepackage{cleveref}

\begin{document}
\pagestyle{plain} 
\title{Contract Usage and Evolution in Android Mobile Applications}


\author{David R. Ferreira}
\affiliation{%
 \institution{Faculty of Engineering, University of Porto}
  \city{Porto}
  \country{Portugal}
}
\email{david.regatia@gmail.com}

\author{Alexandra Mendes}
\affiliation{%
 \institution{HASLab / INESC TEC \& Faculty of Engineering, University of Porto}
  \city{Porto}
  \country{Portugal}
}
\email{alexandra@archimendes.com}

\author{Jo\~{a}o F. Ferreira}
\affiliation{%
  \institution{INESC-ID \& IST, University of Lisbon}
  \city{Lisbon}
  \country{Portugal}
}
\email{joao@joaoff.com}

\renewcommand{\shortauthors}{Ferreira, Mendes, and Ferreira}

\begin{abstract}

Formal contracts and assertions are effective methods to enhance software quality by enforcing preconditions, postconditions, and invariants. Previous research has demonstrated the value of contracts in traditional software development contexts. However, the adoption and impact of contracts in the context of mobile application development, particularly of Android applications, remain unexplored.

To address this, we present the first large-scale empirical study on the presence and use of contracts in Android applications, written in Java or Kotlin. We consider different types of contract elements divided into five categories: conditional runtime exceptions, APIs, annotations, assertions, and other.
%
We analyzed 2,390 Android applications from the F-Droid repository and processed more than 51,749 KLOC
to determine
1) how and to what extent contracts are used, 2) how contract usage evolves, and 3) whether contracts are used safely in the context of program evolution and inheritance. 
%
Our findings include: 1) although most applications do not specify contracts, annotation-based approaches are the most popular among practitioners; 2) applications that use contracts continue to use them in later versions, but the number of methods increases at a higher rate than the number of contracts;
and 3) there are many potentially unsafe specification changes when applications evolve and in subtyping relationships, which indicates a lack of specification stability.
Our findings show that it would be desirable to have libraries that
standardize contract specifications in Java and Kotlin, and tools that aid practitioners in writing stronger contracts and in detecting contract violations in the context of
program evolution and inheritance.


\end{abstract}

\ccsdesc[500]{General and reference~Empirical studies}
\ccsdesc[500]{Software and its engineering~Software evolution}
\ccsdesc[300]{General and reference~Reliability}

\keywords{design by contract, software reliability, android applications, contracts, assertions, verification, Kotlin, Java, preconditions, postconditions, invariants}



\maketitle

\section{Introduction}

Building reliable mobile applications is a growing concern as many companies are adopting iOS and Android as target platforms for their apps in critical domains such as mobility, health, finance, and government. 
There are now more mobile phones than people in the world: in 2022, there were more than 8.58 billion mobile subscriptions in use worldwide,\footnote{\url{https://www.weforum.org/agenda/2023/04/charted-there-are-more-phones-than-people-in-the-world/} (accessed 17 November 2023)} with more than 2 million apps available on the App Store and Google Play \cite{kn:TaoEdmunds18}. Additionally, data from 2022 shows that Android represents approximately 43\% of the overall operative system market share, followed by Windows (29\%), and then iOS (18\%) \citep{kn:OSUsage2023}. Therefore, faults in mobile apps, and particularly in Android apps, can impact a very large number of users. In addition, with an increasing number of apps in critical areas such as health and finance, 
faults can have a huge negative impact.
It is thus important to use software reliability techniques when developing mobile applications. 


One of these techniques is Design by Contract (DbC) \cite{kn:Meyer1992}, under which software systems are seen as components that interact amongst themselves based on precisely defined specifications of client-supplier obligations (\emph{contracts}). Contracts are specifications of an agreement between the client and the supplier of a component, where the supplier expects that certain conditions are met by the client before using the component (\emph{preconditions}), maintains certain properties from entry to the component to exit (\emph{invariants)}, and guarantees that certain properties are met upon exit (\emph{postconditions}). These contracts are written as \emph{assertions} directly into the code. 
Currently, there are assertion capabilities in most programming languages, but assertions are not universally used. 
%
%
Since the 1980s, many have advocated DbC \cite{meyer1988programming} as an efficient technique to aid the identification of failures \citep{kn:Aniche2022}, improve code understanding \citep{kn:Fairbanks19}, and improve testing efforts \citep{kn:Tantivongsathaporn2006} which directly or indirectly contribute to
more reliable software.
This has led to a number of empirical studies on the use of contracts in a variety of contexts \cite{kn:chalin2006,kn:schiller2014,kn:Estler2014,kn:casalnuovo2015,kn:kudrjavets2006,kn:kochar2017,kn:counsell2017,kn:dietrich17}.
However, \emph{there are no previous studies on the presence and usage of contracts in Android applications nor any study that includes the Kotlin language}. 

In this paper, we present the first large-scale empirical study of contract usage in Android mobile applications written in Java or Kotlin. Our goal is to understand what is
the adoption rate of contracts among Android developers, their evolution, and which language features are used to denote contracts. 
Information on how practitioners use contracts can aid the 
creation and improvement of tools and libraries by researchers and tool builders \citep{kn:schiller2014}. Additionally, empirical evidence about the benefits of contracts can encourage their adoption by practitioners and the establishment of DbC as a software design standard \citep{kn:Tantivongsathaporn2006}.

We aim to determine: 1) how and to what extent contracts are used, 2) how contract usage evolves in an application, and 3) whether contracts are used safely in the context of program evolution and inheritance. 



\newpage
\noindent In summary, the contributions of this paper are:
\begin{itemize}[leftmargin=1.9em]
    \item The first large-scale empirical study about contract usage and evolution in Android applications, resulting in a list of findings and recommendations for practitioners, researchers, and tool builders. 
    Also, no previous studies consider Kotlin programs.
    \item A list of language features, tools, and libraries to represent contracts in Android applications.
    \item A dataset of 1,767 Java and 623 Kotlin Android applications, together with a pipeline of scripts that can be used to build large-scale datasets of Java and Kotlin open-source projects, guided by inclusion criteria and size optimization. 
    \item An updated and extended version of Dietrich et al.'s tool~\cite{kn:dietrich17}, which can now analyze Kotlin code and can be used to investigate additional Android-specific contracts.
\end{itemize}
It should be noted that, even though we update and extend \citet{kn:dietrich17}'s tool, our work \emph{is not} a replication of their study. Our study differs from their work by focusing on Android applications and not on Java applications only. Due to the focus on Android, our study considers Kotlin in addition to Java, 
as since 2019, the
Kotlin programming language is the preferred language for Android app developers\footnote{\url{https://techcrunch.com/2019/05/07/kotlin-is-now-googles-preferred-language-for-android-app-development} (accessed 17 November 2023)}.
Moreover, Kotlin is now used by over 60\% of Android professional developers\footnote{\url{https://developer.android.com/kotlin} (accessed 17 November 2023)}.

\paragraph{\textbf{Data \& Artifact Availability}}
To support our study, an artifact was developed to automatically collect contracts from Android applications and to produce the necessary empirical data. The artifact is written in Python and Java, and includes an extension of the tool proposed by Dietrich et al.~\cite{kn:dietrich17}. 
All the code and datasets are publicly available: 
\url{https://figshare.com/s/d6eb7e5522bb535dc81a} 

\section{Contracts in Android Applications}
\label{sec:contracts-android-apps}

\setlength{\textfloatsep}{0.5cm}

\begin{table}
\centering
\footnotesize
\caption{Contract elements considered in this study}
\label{tab:contracts-considered}
\begin{tabular}{@{}ll@{}} \toprule
   \textbf{category} & \textbf{examples} \\ \midrule
   CREs                & \mintinline{java}{IllegalArgumentException}\\
    (74 constructs) & \mintinline{java}{EmptyStackException} \\
   ~& \mintinline{java}{SecurityException} \\
   ~& \mintinline{java}{UnsupportedOperationException} \\
   ~& \mintinline{java}{AccessControlException} \\
   ~& \mintinline{java}{IndexOutOfBoundsException}\\ 
   ~& \mintinline{java}{NullPointerException}\\ 
                       \midrule
   APIs                & \mintinline{java}{org.apache.commons.lang.Validate.*}  \\
   (31 constructs) & \mintinline{java}{org.apache.commons.lang3.Validate.*}  \\
                       & \mintinline{java}{com.google.common.base.Preconditions.*}  \\
                       & \mintinline{java}{org.springframework.util.Assert.*}  \\
                       \midrule
   Assertions          & \mintinline{java}{assert} (Java) \\ 
   (6 constructs)      & \mintinline{java}{assert} (Kotlin) \\
                       & \mintinline{java}{check(), checkNotNull()} (Kotlin) \\
                       & \mintinline{java}{require(), requireNotNull()} (Kotlin) \\ 
                       \midrule
Annotations         & \mintinline{java}{org.jetbrains.annotations.*}  \\
   (136 constructs) & \mintinline{java}{org.intellij.lang.annotations.*}  \\
                       & \mintinline{java}{edu.umd.cs.findbugs.annotations.*}  \\
                       & \mintinline{java}{android.annotation.*}  \\
                       & \mintinline{java}{androidx.annotation.*}  \\
                       & \mintinline{java}{javax.annotation.*} (JSR305) \\
                       \midrule
   Other               & \mintinline{kotlin}{@ExperimentalContracts} (Kotlin) \\ 
   (1 construct)       & \\
                       \bottomrule
\end{tabular}
\end{table}
Our notion of contract follows from the theory of \emph{design by contract}~\cite{kn:Meyer1992}, where
preconditions, postconditions, and invariants are used to document (and specify) state changes that might occur in a program. Preconditions and postconditions are associated with individual methods and constrain their input and output values. On the other hand, invariants are associated with classes and properties and constrain all the public methods in a given class.  Preconditions represent the expectations of the contract, and postconditions represent its guarantees. Invariants represent the conditions that the contract maintains.

Contrary to the Eiffel language, conceived by Bertrand Meyer in 1985, neither Java nor Kotlin provide a native and standardized approach for contract specification \citep{kn:chalin2006}. Still, developers can take advantage of language features and libraries to specify preconditions, postconditions, and class invariants in both languages. For example, they can use constructs provided by the programming language, such as the Java \mintinline{java}{assert} keyword introduced in Java 1.4; they can use conditional runtime exceptions such as Java \mintinline{java}{IllegalArgumentException}; they can use annotations such as the AndroidX annotations \mintinline{java}{@NonNull} and \mintinline{java}{@Nullable}; and they can use specialized libraries such as Google Guava's Preconditions API.\footnote{\url{https://guava.dev/releases/snapshot-jre/api/docs/com/google/common/base/Preconditions.html} (accessed 31 July 2023)}

To facilitate the comparison with previous studies,
we group these constructs into the five categories proposed by Dietrich et al.~\cite{kn:dietrich17}:
conditional runtime exceptions (CREs), APIs, annotations, assertions, and other. The main difference is that, since we focus on Android applications, we include contract elements that are specifically used by Android developers (e.g., Android annotations and specific Android runtime exceptions).
Table~\ref{tab:contracts-considered} summarizes the classification and provides some examples; we consider a total of 248 constructs. 
Below, we briefly describe each category.

\paragraph{\bfseries CREs} 

An exception can be used to signal, at runtime, a contract violation. Bloch~\cite{bloch2008effective} suggests the use of runtime exceptions to indicate
programming errors, as the great majority indicates precondition violations. However, it is important to note that the exception itself does not represent a contract; it needs to be associated with a previous check --- for example, an exception thrown inside an \emph{if-else block} --- to be considered so. Java and Kotlin offer many exceptions that can be used for this purpose, such as the \emph{IllegalArgumentException}. The \emph{android.util} package offers additional exceptions that we are also interested in analyzing, such as the case of the \emph{AndroidRuntimeException}. 
We are also interested in a particular exception, the \textit{UnsupportedOperationException}, which, according to the Java documentation, is thrown to indicate that the requested operation is not supported. As \citet{kn:dietrich17} argues, this is the strongest possible precondition and can not be satisfied by any client. 

%
%
%
When compared to \citet{kn:dietrich17}, our list of CREs includes 74 exceptions versus 6 exceptions used in their study.
We show some examples in Table~\ref{tab:contracts-considered} but, due to the paper length restriction, these are fully listed in the Supplementary Material. 

\paragraph{\bfseries Annotations} Annotations are metadata added to the program providing information that can be used at compile time or runtime to perform further actions. Java provides many annotations through the \textit{java.lang} package. \autoref{tab:contracts-considered} lists the annotation packages we are particularly interested in studying in the context of contracts. 
No previous studies consider the
\emph{android.annotation} and the \emph{androidx.annotation} packages. 

The annotation-based approach is particularly interesting due to two reasons. Firstly, many annotations can be associated with the method's arguments (preconditions), the method's return values (postconditions), or the class properties (invariants). Secondly, since annotations are usually added to the method's signature or to the class property, there is a greater separation between the contract specification and the service's implementation. This means that annotations, like in the Eiffel's approach, do not increase the complexity of the method's implementation, contrary to what happens with CREs, APIs, and assertion-based approaches.

\paragraph{\bfseries APIs}\label{paragraph:APIs} APIs consist of wrappers around conditional exceptions and other basic constructs. This contributes to a simpler, more verbose, and explicit representation of contracts. We are interested in the four APIs listed in \Cref{tab:contracts-considered}. 
For example, the \emph{Apache Commons} offers the \emph{Validate}\footnote{\url{https://commons.apache.org/proper/commons-lang/apidocs/org/apache/commons/lang3/Validate.html} (accessed 4 June 2023)} class that according to the official documentation ``assists in validating arguments'', which suggests a precondition usage. The same documentation also states that the validation methods follow the following principles: a null argument leads to a \emph{NullPointerException}; a non-null argument leads to an \emph{IllegalArgumentException}; an index issue in a collection-type structure leads to an \emph{IndexOutOfBoundsException}.

The methods provided by the \emph{Validate} class are simply wrapping exceptions that we have already considered in the CREs.
Interestingly, as we can perceive from these APIs and their documentation, they are built to be used as preconditions. The same libraries do not offer any equal approach to specify postconditions which 
shows a preference from tool builders into preconditions over postconditions. Nevertheless, and against the documentation guidelines, practitioners can still use any of those API's methods to check postconditions.

\paragraph{\bfseries Assertions} Assertions have been introduced in Java 1.4 and are specified through the \emph{assert} reserved keyword. It helps practitioners verify conditions that must be true during runtime. JVM throws an \emph{AssertionError} if the condition is false. However, JVM disables assertion validation by default, requiring it to be explicitly enabled. This means that the practitioner may be assuming that the contracts specified through assertions will be validated at runtime when in fact the assertions are disabled. This leads to an incorrect, and potentially dangerous, assumption. Having that in mind, assertions can still easily be used to check preconditions and postconditions. 

Kotlin also has its own \emph{assert}. However, contrary to the Java version, \emph{assert} in Kotlin is a function and not a reserved word. This means that any class can define a method with the name \emph{assert}, which makes it harder for an automated analysis tool to distinguish between Kotlin's assert or a developer's custom method that does something else. Additionally, contrary to Java, Kotlin always executes the assert expression and only uses the \emph{-ea} JVM flag to decide whether to throw the exception, which can cause problems on performance-sensitive applications.
Additionally, Kotlin offers other methods --- \emph{check()}, \emph{checkNotNull()}, \emph{require()}, and \emph{requireNotNull()}. Although these throw an \emph{IllegalArgumentException} or an \emph{IllegalStateException} instead of an \emph{AssertionError}, we added them to the assertions category because of their syntactic similarities. 

\paragraph{\bfseries Other} 
In our study, we consider Kotlin Contracts\footnote{\url{https://github.com/Kotlin/KEEP/blob/master/proposals/kotlin-contracts.md} (accessed 31 July 2023)}, 
an experimental feature introduced in Kotlin 1.3 that allows the developer to state a method's behavior to the compiler explicitly.
Kotlin contracts provide a way to explicitly express aspects of function's behavior, thus qualifying as a construct of interest to our study. As the following example shows, they also provide useful information to the compiler: the call to \emph{split} in line~\ref{code:split} causes no error, because the contract specified in line~\ref{code:contract} guarantees that \emph{birthdate} is not null.


\begin{lstlisting}[
language=Java, 
label={lst:3-others-kotlinContracts2},
aboveskip=1em,
belowskip=1em,
frame=none,
backgroundcolor=\color{white}
]    
  @ExperimentalContracts
  fun sendBirthdayMessage(birthdate: String?) {
    isBirthdateValidOrElseThrow(birthdate)
    val birthdaySplit = birthdate.split("/") // no error %*\label{code:split}*)
    ...
  }

  @ExperimentalContracts
  fun isBirthdateValidOrElseThrow(birthdate: String?) {
    contract { returns() implies (birthdate != null) } %*\label{code:contract}*)
      if (birthdate == null) {
        throw IllegalArgumentException()
      } ...
  }
\end{lstlisting}

\section{Related Work}

This section presents related work on the usage of contracts, assertions, and annotations by practitioners.

\paragraph{\textbf{DbC and Contracts Usage}} \label{sec:empiricalStudiesContracts}
It is widely supported that DbC contributes to improve software reliability \citep{kn:murthy2019, kn:furiaWeiKazmin2011, kn:Hollunder2012}. The advantages commonly mentioned by various authors are that DbC (i) improves code understanding \citep{kn:Fairbanks19, kn:naumchev2019, kn:furiaWeiKazmin2011, kn:silva2020}, (ii) helps identify bugs earlier and diagnose the failure \citep{kn:furiaWeiKazmin2011, kn:Aniche2022, kn:casalnuovo2015, kn:dietrich17, kn:schiller2014}, and (iii) contributes to better tests \citep{kn:furiaWeiKazmin2011, kn:Aniche2022, kn:schiller2014, kn:algarni2018, kn:Tantivongsathaporn2006}.
%
%
Some studies 
demonstrated that DbC requires fewer project person-to-hour resources \cite{kn:Blom2002,kn:Tantivongsathaporn2006}, but could not confirm an impact on quality. 
Moreover, DbC contributes to less time spent on writing tests~\cite{kn:Tantivongsathaporn2006}.
Blom et al. \cite{kn:Blom20022} presented a DbC-based development strategy and a case study applying it to an enterprise project. They found that the approach resulted in fewer errors and decreased development time. In another study, 
\citet{kn:Zhou2017} concluded that DbC increased reliability in software components. 
In a study on C\# projects using Code Contracts, \citet{kn:schiller2014} found a high percentage of contracts related to null checking
and suggest the importance of creating design patterns alongside tools and libraries. 
\citet{kn:Estler2014} analyzed a dataset of 21 Eiffel, C\#, and Java projects 
known to be equipped with contracts. Most contracts are null checks, with preconditions being typically larger than postconditions. The authors concluded that the average number of clauses per specification is stable over time and that the method's implementation changes more frequently than its specification. However, they warned that strengthening contracts may be more frequent than weakening, indicating some unsafe evolution of contracts.
Lastly, \citet{kn:dietrich17} investigated 176 popular Java projects 
in the Maven repository and found that the majority of programs do not use contracts significantly. 
They found that CREs are the most commonly used category, followed by asserts. The dominance of preconditions over postconditions in contracts is consistent with other studies \citep{kn:chalin2006,kn:schiller2014}. Projects that use contracts maintain or even expand their usage over time, according to the authors' analysis of contract usage through program evolution. Similarly to \citet{kn:Estler2014}, the authors also reported some unsafe evolution of contracts, which can happen when a method strengthens its preconditions
or weakens its postconditions.
They also found many violations of the Liskov Substitute Principle, with prevalence in the annotations. According to this principle, a sub-type can only weaken preconditions or strengthen postconditions and class-invariants from its parent \citep{kn:feldman2006}. The authors caution that their dataset mainly includes libraries, which may explain the low usage of annotations. This study is the one most related to the work presented here, as it also studies contracts in Java. However, our study differs from \citet{kn:dietrich17} in that we focus on Android applications and we study both Java and Kotlin. Moreover, we consider more constructs.

\paragraph{\textbf{Assertion Usage}} \label{sec:empiricalStudiesAssert}


\citet{kn:kudrjavets2006} studied two Microsoft Corporation components, written mainly in C and C++, and found that increased assert density led to a decrease in faulty density in C and C++ code and that using asserts was more effective for fault detection than some static analysis tools.
%
%
\citet{kn:kochar2017} studied a dataset of 185 Apache Java projects available on GitHub and found that adding asserts contributes to fewer defects, especially when many developers are involved. This agrees with reports from \citet{kn:kudrjavets2006} but is not supported by \citet{kn:counsell2017} that analyzed two industrial Java systems and found no evidence that asserts were related to the number of defects. \citet{kn:kochar2017} also concluded that developers with more ownership and experience use asserts more often, which shows that more advanced programmers see it as a valuable practice. In line with other previously mentioned studies for contracts \citep{kn:schiller2014, kn:Estler2014}, most uses are related to null-checking.



\paragraph{\textbf{Annotation Usage}} \label{sec:empiricalStudiesAnnotations}
There is a general understanding that the use of annotations among practitioners is ever-growing \citep{kn:yu2021, kn:grazia2022}. 
\citet{kn:yu2021} conducted a study on 1,094 GitHub open-source projects
and found a median value of 1.707 annotations per project, 
with some developers overusing them. 
The authors argue the need for better training and tools to help derive better annotations. Other authors made a similar claim for contracts \citep{kn:schiller2014}. Additionally, developers with higher ownership use annotations more often, which agrees with the findings by \citet{kn:kochar2017} related to assertion usage.
\citet{kn:grazia2022} investigated the evolution of type annotations, some of which can act as contracts, in 9,655 Python prototypes. 
The authors reported that although type annotations usage is increasing, less than 10\% of potential elements are being annotated. This contradicts the (general) annotations overuse reported by \citet{kn:yu2021}. More importantly, the study found that once added, 90.1\% of type annotations are never updated. This indicates that specifications are more stable than implementations, which is desirable. A similar finding was reported by \citet{kn:Estler2014} related to the stability of contracts while the program evolves. Also relevant is that most type annotations were associated with parameter and return types, rather than with variable types. 
Finally, the authors found that adding type annotations increased the number of detected type errors. This motivates the general use of these features to improve software reliability.

%

\section{Study Design} 
    In this section, we present the design of our study, including
    the research questions, how the dataset of 
    Android applications is created, the classification used for contracts,
    and the methodologies used to study contract usage and evolution.

\subsection{Research Questions}
\label{sec:research-questions}
    
    In this study, we aim to answer the following research questions:

    \begin{itemize}
    \item[\textbf{RQ1}.] {\textbf{[Contract Usage]}} How and to what extent are contracts used in Android applications?
    \item[\textbf{RQ2}.] {\textbf{[Evolution]}} How does contract usage evolve in an application?
    \item[\textbf{RQ3}.] {\textbf{[Safety]}} Are contracts used safely in the context of program evolution and inheritance?
    \end{itemize}

\subsection{Dataset}


The dataset used is composed of real-world applications obtained from F-droid,\footnote{\url{https://f-droid.org} (accessed 17 November 2023)} an alternative app store listing over 4,000 free and open-source projects. The fact that it has a large number of open-source apps (with public source code) on a wide range of domains, makes F-Droid a good option. Moreover, F-Droid is normally used in research studies on Android applications~\cite{chen2019storydroid,zeng2019studying}. Apart from native Android applications written in Java or Kotlin, F-Droid's catalog also contains projects that use hybrid frameworks (e.g., React Native) that we exclude from our dataset.

We started by downloading the \emph{F-Droid index}, which is a list of URLs for each project available in the catalog. Next, this list is \emph{filtered} based on the following criteria:
\begin{enumerate*}
    \item The application source code is hosted in GitHub;
    \item The application source code is either Java or Kotlin;
    \item The GitHub project is not archived;
    \item The GitHub project has had a commit since 2018.
\end{enumerate*}
These inclusion criteria ensure that the project's source code is easily accessible (through GitHub), is written mainly in Java or Kotlin (the languages we are interested in studying), while also guaranteeing that the project is active and relevant. We retrieve \emph{two versions} for each of the filtered projects, which is a required step for the evolution study. We do this by storing a list of the URLs pointing to two GitHub versions: we first try to retrieve the oldest and the most recent \emph{release}; if there are not enough releases, we try to retrieve the oldest and the most recent \emph{tag}; finally, if there are not enough tags, we just keep the most recent commit of the repository. Although our script resolved most of the versioning schemes found, some projects required manual handling to determine which version was the first and the last. In the rest of the paper we also refer to the most recent version as \emph{last} or \emph{second} version.
Finally, we clone all the projects contained in the versions list. \emph{Every file that is neither a Java nor a Kotlin file is removed} from the dataset, which helps to decrease its size. 

\begin{table}[]
\footnotesize
\renewcommand\arraystretch{1.1}
\centering
\caption{Dataset metrics.}
\begin{tabular}[t]{llll}
    \hline
    \hline
    \textbf{metric} & \textbf{Java} & \textbf{Kotlin} & \textbf{Both} \\ 
    \midrule
    projects & 1,767 & 623 & 2,390 \\ 
    compilation units & 204,468 & 123,222 & 327,690 \\
    classes & 299,824 & 251,632 & 551,456 \\
    methods (all) & 2,079,276 & 602,884 & 2,682,160 \\
    constructors (all) & 205,773 & 94,866 & 300,639 \\
    methods (public, protected, internal) & 1,770,972 & 482,972 & 2,253,944 \\
    constructors (public, protected, internal) & 184,868 & 93,587 & 278,455 \\
    KLOC including comments & 40,041 & 11,708 & 51,749 \\
    \bottomrule
\end{tabular}
\label{tab:data-metrics}
\end{table}
\paragraph{\textbf{Dataset metrics.}}
From the initial list of 4,070 projects in the F-Droid index retrieved on May 21, 2023, we got 3,215 hosted in GitHub, 3,141 non-duplicated URLs, and 2,390 projects after filtering by the inclusion criteria. Out of these, 1,767 are Java applications and 623 are Kotlin applications.
Since we tried to retrieve two versions for each application, we analyzed 4,192 \emph{program-version pairs}. This means that for 294 applications, it was only possible to retrieve a single version. While these applications are still evaluated in the context of the \emph{usage} and \emph{LSP} studies, they are not considered for the \emph{evolution} study. 


\autoref{tab:data-metrics} presents additional metrics about the final dataset size used in the empirical evaluation. Additionally, the table shows that the dataset is imbalanced, with more Java applications. The dataset includes 204,468 Java and 123,222 Kotlin compilation units and, therefore, Java represents 62.4\% of the overall number of compilation units. This imbalance requires caution when trying to read this work's results from the perspective of comparing Java against Kotlin's use of contracts. Furthermore, the dataset includes 551,456 classes,  2,682,160 methods, and 300,639 constructors. Note that we did not consider \emph{private} methods, because those methods are not used directly by a client, and a contract is a bond between a supplier and a client. In total, we analyzed 2,532,399 \emph{public}, \emph{protected}, and \emph{internal} methods, and constructors. 
It is also worth noting the difference in the number of \emph{compilation units} and \emph{classes} between the two languages. Although the dataset contains 1.67 times more Java compilation units than Kotlin ones, it only includes 1.19 times more Java classes. Since a compilation unit usually represents a file, this means that there are more classes per file in Kotlin (2.04) than in Java (1.47). This is expected due to the more restricted Java rules that, for example, only allow a single top-level public class per file.

From the standpoint of the dataset's diversity, it includes apps from various domains, such as gaming, communication, multimedia, security, health, and productivity. 

\subsection{Data Collection and Analysis} 
\label{sec:dataCollectionAnalysis}
Here, we describe the analysis tool used and the three studies that we conducted to answer our research questions: the usage study, the evolution study, and the Liskov Substitution Principle study.

\subsubsection{Analysis Tool}\label{sec:analysisTool}
Our analysis tool is an extension of the tool created by Dietrich et al.~\cite{kn:dietrich17}, which was used in their study on the usage of contracts in Java applications. 
We extended the tool to support Kotlin source code and more constructs focused on Android applications. Additionally, the framework's code also suffered considerable refactoring and organization to simplify and ease its comprehension and maintainability.
The main effort was to add support for Kotlin source code. The original tool was using the JavaParser\footnote{\url{https://javaparser.org} (accessed 17 November 2023)} library to perform AST analysis of Java code. Since this library is not able to parse Kotlin source code, we integrated JetBrains's Kotlin compiler\footnote{\url{https://github.com/JetBrains/kotlin} (accessed 4 June 2023)} to perform this task. This required us to implement new versions of the tool's extractors and visitors classes using the methods provided by the new library to be able to identify contract patterns in Kotlin. We also updated the JavaParser library to support newer Java versions.


The tool is divided into three main parts: 1) \emph{usage}, which extracts the list of contracts present in each program and produces statistics about their use; 2) \emph{inheritance}, which identifies contracts in overridden methods and validates whether they violate the Liskov Substitution Principle; and 3) \emph{evolution}, which analyses how identified contracts evolve in later versions of the application. The following sections describe how each component contributes to answering the proposed research questions.

\subsubsection{Usage Study} \label{sec:usage-study}

The usage study is divided in two main steps: 1) identifying contract occurrences and 2) producing statistics about those results. Our tool uses the JavaParser and JetBrains's Kotlin compiler libraries to perform AST analysis of all dataset's source code file that is either Java or Kotlin. This analysis is done against a set of extractors to identify occurrences of our defined constructs. Each category requires different approaches for their identification:

\begin{itemize}[leftmargin=1.1em]
    \item \emph{CREs}. During the AST analysis, we look for the pattern: 
    \[
    \small
    \texttt{if (<condition>) \{ throw new <exception> (<args>) \}}
    \]
    When this pattern is found, we check whether the exception belongs to the list of CREs considered (see \Cref{sec:contracts-android-apps}). In line with Java's good practices, we assume that CREs are used with preconditions.

    \item \emph{APIs}. Firstly, we check whether the file contains an import declaration to any API package listed in \autoref{tab:contractsbycategory}
    . If any is found, all call expressions in that file are analyzed to determine if they are invoking any of the methods provided by the API. As stated in \Cref{paragraph:APIs}, we assume the analyzed APIs to be associated with preconditions. 

    \item \emph{Assertions}. Identifying Java asserts is straightforward since the JavaParser provides a visitor method for this particular statement. The complexity lies in identifying Kotlin asserts, which is not a reserved keyword. To handle this challenge, when analyzing a file, we first search for any method declaration and any import statement that has a name equal to one of the following expressions: \emph{assert}, \emph{require}, \emph{requireNotNull}, \emph{check}, \emph{checkNotNull}. Next, we identify whether the class invokes any method with one of those names. Suppose a class has a method declaration/import statement and an invocation with one of these expression's names. In that case, we consider it an ambiguous situation, and therefore, we do not consider it an assert instance. If the class invokes one of those methods but does not declare/import any method with that same name, we consider it an assert.  This is not a fool-proof approach, but it minimizes the under-reporting to a residual level. We do not classify assertions either as preconditions or postconditions.

    \item \emph{Annotations}. We check if the source code file contains an import statement to one of the packages listed in \autoref{tab:contracts-considered}. If that is the case, we check every annotation in that file to see if it matches any of those provided by the imported package. We also identify the artifact to which the annotation is associated as follows: 1) annotations associated with a method's parameters are preconditions; 2) annotations associated with a method are postconditions; and 3) annotations associated with a field are class invariants.

    \item \emph{Others}. This category only includes the investigation of the experimental \emph{Kotlin Contracts}. To identify occurrences of this construct, we look for the pattern - \textit{contract \{returns (<condition>) implies (<condition>)\}}.
\end{itemize}
Our tool creates a \textit{JSON} file for each program-version pair that stores the 
identified contracts, including 1) the file path, 2) the associated condition, 3) the method or property name, 4) the type of artifact (method or property), 5) the line number, and 6) the contract type. 
In the second step of the \emph{usage} study, all the \textit{JSON} files are analyzed to produce statistics about the identified contracts, including the frequency of each category (API, annotation, assertion, etc.), class (preconditions, postconditions, and class invariants), construct (java assert, Guava API, \textit{androidx} annotations, etc.), to compute the Gini coefficient for each category, and to list the programs with more contracts for each category.


\subsubsection{Evolution Study} \label{sec:evolution-study}
In the evolution study, we are interested in knowing what happens to contracts while the application evolves. In other words, after identifying a contract in the first version of the application, we check whether, in the later version, the contract still exists, was modified or removed. At the same time, we are also reporting cases when a contract is added to an artifact (method or parameter) in the later version of the app (but was not present in the first version). This information provides insights into how contracts evolve in an application and whether this evolution poses risks to the client.

As already mentioned, a contract establishes rights and obligations for each part --- the client assumes that the supplier will keep its obligations and vice-versa. Therefore, when a contract is altered, both parts should be informed and updated accordingly. This is particularly crucial when a \emph{precondition is strengthened} or when a \emph{postcondition is weakened}. In the first case, if the precondition is strengthened and the client does not know it, it can fail to cover its new obligations, and, therefore, the supplier is not bound to keep its part of the contract. In the latter case, if the postcondition is weakened, the client may still be making assumptions that the supplier does not ensure anymore. 
An example is shown in ~\autoref{lst:3-evolution-evolution-0}, where the annotation \emph{@NonNull} was added to the \emph{toolbar} parameter in the last version. This is the case of a \emph{precondition strengthening}: in the first version, the method accepted a null \emph{toolbar}, but now it requires it to be not null. Therefore, if the client is not updated, it will fail to cover its new obligation.


\definecolor{addedcolor}{RGB}{34, 139, 34}  
\definecolor{removedcolor}{RGB}{220, 20, 60} 
\newcommand{\addline}{\makebox[0pt][l]{\color{addedcolor!30}\rule[-0.45em]
{\linewidth}{1.5em}}}
\newcommand{\removeline}{\makebox[0pt][l]{\color{removedcolor!30}\rule[-0.45em]{\linewidth}{1.5em}}}
\begin{lstlisting}[
language=Java, 
caption={A change in a Java method signature: version 0 specifies three pre-conditions and version 1 specifies four.}, 
captionpos=b,
label={lst:3-evolution-evolution-0},
morekeywords={@NonNull, @Nullable, @ColorInt},
float,floatplacement=H,
belowskip=0pt,
aboveskip=0pt
]
    public static void setToolbarContentColorBasedOnToolbarColor(
        @NonNull Context context,
%*\removeline{\,\,-\ \ \ \ \ \ \ \ Toolbar toolbar,}*)
%*\addline{\,\,+\ \ \ \ \ \ \ \ \textbf{@NonNull} Toolbar toolbar,}*)
        @Nullable Menu menu,
        int toolbarColor,
        final @ColorInt int menuWidgetColor
\end{lstlisting}


To conduct this study, we follow the same approach as Dietrich et al.~\cite{kn:dietrich17}, firstly creating \textit{diff records} from the contracts present at the two versions of a program's method and then classifying them according to the \emph{evolution patterns} listed in \autoref{table:3-evolution-categories}.


\begin{table}
\footnotesize
\renewcommand\arraystretch{1.1}
\centering
\caption{Classification of the diff records produced during the evolution and LSP study.} 
\begin{tabular}[t]{l>{\raggedright}p{0.45\linewidth}>{\raggedright\arraybackslash}p{0.1\linewidth}}
    \toprule
    \textbf{Classification} & \textbf{Description} & \textbf{Risk} \\ 
    \midrule
    PreconditionsStrengthened & A precondition was added to a method or a clause to an existing precondition with the '\&' or '\&\&' operators. & Potential risk \\
    \hline
    PreconditionsWeakened & A precondition was removed from a method, or a clause was added to an existing precondition with the '|' or '||' operators. & No risk.\\
    \hline
    PostconditionsStrengthened & A postcondition was added to a method or a clause to an existing postcondition with the '\&' or '\&\&' operators. & No risk.\\
    \hline
    PostconditionsWeakened & A postcondition was removed from a method, or a clause was added to an existing postcondition with the '|' or '||' operators. & Potential risk.\\
    \bottomrule
\end{tabular}
\label{table:3-evolution-categories}
\end{table}

\subsubsection{Liskov Substitution Principle Study} \label{sec:lsp-study}

When a method is overridden in a subclass, that class can specify new contracts added to the ones inherited from the superclass method. In this case, proper handling of contracts should follow the Liskov Substitution Principle (LSP), which states that the subclass method must accept all input that is valid to the superclass method and meet all guarantees made by the superclass method. In other words, a subclass method can only \emph{weaken preconditions} and \emph{strengthen postconditions}.

To detect those occurrences, the analysis tool first lists all methods in each program-version pair associated with their respective class. Additionally, it also identifies the class' parents.
%
%
%
Then, similarly to the \emph{evolution} study, diff records are created between the subclass and the superclass methods. These records are classified according to the evolution patterns described in \autoref{table:3-evolution-categories}.


\section{Results}
In this section, we present
the results of our empirical study, as well as the main findings.

\subsection{RQ1: Contract Usage} 

%
\autoref{tab:contractsbycategory} shows the number of contracts found per category, considering all versions (columns 2 and 3) and considering only the latest version of each application (columns 4 and 5). The table also identifies the number of applications containing at least one contract for that category (columns 6 and 7). The most obvious conclusion is that, in both languages, annotation-based contracts are the most popular category. More specifically, considering both languages in the last version, annotations represent 87.1\% of the contracts found, followed by CRE with 9.7\%, and then assertions with 2.5\%. The results show similar tendencies between Java and Kotlin, and the only difference is that while Java's second most popular category is CREs, in Kotlin, it is assertions. This relatively high percentage of the assertion category in Kotlin is explained by our inclusion of the four language's standard library methods listed in \Cref{sec:contracts-android-apps}, where \emph{require()} alone counts 901 total occurrences distributed by 112 last versions.

\begin{tcolorbox}[
    enhanced jigsaw,
    sharp corners,
    boxrule=0.5pt, 
    colback=blue!5!white,   
    boxrule=0pt, 
    frame empty
 ]
\textbf{Finding 1:} Most contracts are annotation-based, accounting for 88.31\% in Java and 77.44\% in Kotlin of the total number of contracts found. 
\end{tcolorbox}

This distribution in categories' popularity significantly differs from the findings of Dietrich et al.~\cite{kn:dietrich17}, who reported that the most common category was CREs and found surprisingly low use of annotations. This may be explained by the difference between the datasets' nature. While our dataset is formed mostly by user-focused Android applications, the author's dataset was mainly Java libraries. In \autoref{tab:contractsbytype}, we can also see that most annotations found belong to the \textit{androidx.annotation.*} package that the authors did not consider since it is Android-specific. Nevertheless, the high number of annotation-based contracts found is in line with literature that supports its increasing popularity \citep{kn:yu2021, kn:grazia2022}. 

From \autoref{tab:contractsbycategory}, we also verify that the usage of \emph{APIs} is very low in both languages, and it is even more residual in Kotlin applications, where only nine instances were found in the latest versions. 
The known industry skepticism around adding third-party dependencies to projects, which may lead to maintainability and support issues in the future, may explain this finding \citep{kn:backes2016, kn:wang2020}.

\begin{tcolorbox}[
    enhanced jigsaw,
    sharp corners,
    boxrule=0.5pt, 
    colback=blue!5!white,   
    boxrule=0pt, 
    frame empty
 ]
\textbf{Finding 2:} The use of APIs to specify contracts is very rare.
\end{tcolorbox}

\begin{table}[]
\footnotesize
\renewcommand\arraystretch{1.1}
\centering
\caption{Number of contracts found in the dataset by category.}
\begin{tabular}[t]{ccccccc}
    \hline
    \hline
     & \multicolumn{2}{c}{contracts (all ver.)} & \multicolumn{2}{c}{contracts (2nd ver.)} & \multicolumn{2}{c}{applications} \\ 
\cmidrule(lr){2-3} \cmidrule(lr){4-5} \cmidrule(lr){6-7}
Category & Java & Kotlin & Java & Kotlin & Java & Kotlin \\
    \midrule
    API & 1,813 & 10 & 1,121 & 9 & 24 & 4 \\ 
	annotation & 158,400 & 24,125 & 139,507 & 15,068 & 1,097 & 541 \\ 
	assertion & 3,525 & 3,746 & 2,186 & 2,239 & 326 & 232 \\ 
	CRE & 26,061 & 3,292 & 15,150 & 2,139 & 789 & 287 \\ 
	other & - & 1 & - & 1 & - & 1 \\
    \bottomrule
\end{tabular}
\label{tab:contractsbycategory}
\end{table}

In \autoref{tab:contractsbytype}, we have a more detailed perspective by having the frequency of each construct. Firstly, we again highlight that the high number of annotations found is leveraged mostly by the \emph{androidx.annotation.*} package. In APIs, the \emph{Guava} library constitutes most of the usage. We were not expecting to see any usage of \emph{Spring Framework Asserts} since this library was designed to be used in the \emph{Spring} framework, but we still found one occurrence. At the same time, we found no occurrences of the now deprecated \emph{FindBugs} annotations. Additionally, we identified a single occurrence of \emph{Kotlin Contracts}, which may depict the practitioner's distrust of using a feature still in an experimental phase.

We now consider 
Table~\ref{tab:contractsGini},
which presents each category's computed \emph{Gini coefficient}. The \emph{Gini coefficient} measures the inequality among the values of a frequency distribution. 
\setlength{\intextsep}{0pt}%
\setlength{\columnsep}{1em}%
\begin{wrapfigure}{r}{0.4\columnwidth}
\makeatletter
\def\@captype{table}
\makeatother
\footnotesize
\renewcommand\arraystretch{1.1}
\centering
\caption{Gini coefficient by category.}
\begin{tabular}[t]{ccc}
    \hline
    \hline
    \textbf{Category} & \textbf{Java} & \textbf{Kotlin} \\ 
    \midrule
	assertion & 0.70 & 0.71 \\ 
	API & 0.80 & 0.37 \\ 
	annotation & 0.88 & 0.76 \\ 
	CRE & 0.77 & 0.67 \\ 
	others & - & 1.00 \\ 
    \bottomrule
\end{tabular}
\label{tab:contractsGini}
\end{wrapfigure}
In other words, a \emph{Gini coefficient} of 0 indicates perfect equality, where all applications have the same number of contracts. In contrast, a \emph{Gini coefficient} of 1 means that a single program has all the contracts. 
We observe that all coefficients in the table are higher than 0.50, except for Kotlin's API usage. The fact that almost all coefficients are very high (close to 1) means that although some applications use contracts intensively, the majority does not use them significantly. This aligns with the results found by Dietrich et al.~\cite{kn:dietrich17}. This conclusion can also be seen in \autoref{tab:topusers}, where the five projects that use more contracts per category are listed. We find that a small group of projects own a large percentage of the overall use in each category. Additionally, it is clearly visible from the \emph{assertion} and \emph{CRE} categories that the numbers quickly decrease through the first to the fifth application showing the unbalanced usage between applications.

\begin{tcolorbox}[
    enhanced jigsaw,
    sharp corners,
    boxrule=0.5pt, 
    colback=blue!5!white,   
    boxrule=0pt, 
    frame empty
 ]
\textbf{Finding 3:} Although there are some applications that use contracts intensively, the majority do not use them significantly.
\end{tcolorbox}

\begin{table}[]
\footnotesize
\renewcommand\arraystretch{1.1}
\centering
\caption{Number of contracts found in the dataset by construct and category.}
\begin{tabular}[t]{cccccc}
    \hline
    \hline
     & & \multicolumn{2}{c}{contracts (all ver.)} & \multicolumn{2}{c}{contracts (2nd ver.)} \\ 
     \cmidrule(lr){3-4} \cmidrule(lr){5-6}
    Construct & Category & Java & Kotlin & Java & Kotlin \\
    \midrule
    cond. runtime exc. & CRE & 25,565 & 3,232 & 14,887 & 2,071 \\ 
    unsupp. op. exc. & CRE & 511 & 142 & 308 & 116 \\ 
    java assert & assertion & 3,525 & - & 2,217 & - \\ 
    kotlin assert & assertion & - & 3,868 & - & 2,370  \\ 
    guava precond. & API & 1,798 & 10 & 1,121 & 9  \\ 
    commons validate & API & 148 & 0 & 3 & 0 \\ 
    spring assert & API & 1 & 0 & 1 & 0 \\ 
    JSR303, JSR349 & annotation & 0 & 0 & 0 & 0 \\ 
    JSR305 & annotation & 4,195 & 20 & 2,133 & 13 \\ 
    findbugs & annotation & 0 & 0 & 0 & 0 \\ 
    jetbrains & annotation & 2,310 & 138 & 1,596 & 98 \\ 
    android & annotation & 12,003 & 5,704 & 7,013 & 3,414  \\ 
    androidx & annotation & 139,933 & 20,593 & 86,212 & 13,811 \\ 
    kotlin contracts & others & - & 1 & - & 1 \\
    \bottomrule
\end{tabular}
\label{tab:contractsbytype}
\end{table}
\begin{table}[]
\footnotesize
\renewcommand\arraystretch{1.1}
\centering
\caption{Top five applications using contracts (second versions only) by category.}
\begin{tabular}[t]{p{0.1\columnwidth}p{0.8\columnwidth}}
    \hline
    \hline
    \textbf{Category} & \textbf{Applications} \\ 
    \midrule
    assertion & K1rakishou-Kuroba-Experimental (378), a-pavlov-jed2k (314), abhijitvalluri-fitnotifications (143), thundernest-k-9 (114), mozilla-mobile-firefox-android-klar (95) \\ 
    CRE & redfish64-TinyTravelTracker (1,036), nikita36078-J2ME-Loader (690), abhijitvalluri-fitnotifications (561), lz233-unvcode-android (561), cmeng-git-atalk-android (447) \\ 
    API & wbaumann-SmartReceiptsLibrary (534), alexcustos-linkasanote (318), BrandroidTools-OpenExplorer (69), oshepherd-Impeller (33), MovingBlocks-DestinationSol (30), inputmice-lttrs-android (24) \\ 
    annotation & MuntashirAkon-AppManager (5,957), Forkgram-TelegramAndroid (5,552), Telegram-FOSS-Team-Telegram-FOSS (5,549), MarcusWolschon-osmeditor4android (4,393), NekoX-Dev-NekoX (4,032) \\ 
    other & zhanghai-MaterialFiles (1) \\ 
    \bottomrule
\end{tabular}
\label{tab:topusers}
\end{table}

Lastly, \autoref{tab:contractsbyclassification} presents the frequency of each contract type. Once again, we have distinct results for Java and Kotlin. In Java, we found 63.73\% of the classified instances in the last versions to be preconditions, 23.19\% postconditions, and only 13.08\% class invariants. These results align with other studies on contracts \citep{kn:chalin2006, kn:schiller2014, kn:dietrich17} that show a clear preference towards preconditions. However, Kotlin's results are different from this expected preference hierarchy. From the classified instances of the 
last versions, we found 38.82\% to be postconditions, 31.73\% class invariants, and 29.44\% preconditions. This suggests that Kotlin developers tend to favor postconditions over any other type, while preconditions come at the last position. 
Also, according to the classification described in \Cref{sec:usage-study}, in our study, only annotations may be classified as postconditions or class-invariants. This means that in Kotlin, there is a higher number of annotations associated with the method's return values and class properties than with the method's parameters. 

\begin{tcolorbox}[
    enhanced jigsaw,
    sharp corners,
    boxrule=0.5pt, 
    colback=blue!5!white,   
    boxrule=0pt, 
    frame empty
 ]
\textbf{Finding 4:} Java and Kotlin practitioners display different tendencies when it comes to the contract type. In Java, there is a clear preference towards preconditions, while in Kotlin, postconditions are the most frequent type.
\end{tcolorbox}

\begin{table}[]
\footnotesize
\renewcommand\arraystretch{1.1}
\centering
\caption{Number of contracts found in the dataset by type.}
\begin{tabular}[t]{ccccccc}
    \hline
    \hline
     & \multicolumn{2}{c}{contracts (all ver.)} & \multicolumn{2}{c}{contracts (2nd ver.)} & \multicolumn{2}{c}{applications} \\ 
     \cmidrule(lr){2-3} \cmidrule(lr){4-5} \cmidrule(lr){6-7}
    Type & Java & Kotlin & Java & Kotlin & Java & Kotlin \\
    \midrule
    pre-condition & 120,671 & 9,203 & 72,160 & 5,744 & 989 & 349 \\ 
    post-condition & 41,764 & 11,490 & 26,253 & 7,575 & 829 & 435 \\ 
    invariants & 23,836 & 9,122 & 14,811 & 6,190 & 643 & 348 \\ 
    not classified & 3,584 & 3,893 & 2,267 & 2,394 & 279 & 202 \\ 
    \bottomrule
\end{tabular}
\label{tab:contractsbyclassification}
\end{table}

Although we can not provide a reason for this finding with certainty, we argue that the difference in practitioners' preference for each type reported in \autoref{tab:contractsbyclassification} could stem from different behavior patterns which are demonstrated in \autoref{tab:popularConstructsPerType-java} and \autoref{tab:popularConstructsPerType-kotlin}. These tables list the ten most occurring constructs for each type in the last versions of Java and Kotlin applications. To create these tables, we followed the classification described in \Cref{sec:usage-study}; hence, for example, although there are 2,217 instances of \emph{JavaAssert} in Java, these were not included in the list since the analysis tool does not classify asserts by type.

By comparing the two tables, we draw distinct behavior patterns between the two languages. In the Kotlin constructs reported by \autoref{tab:popularConstructsPerType-kotlin}, none of the top ten most popular constructs relates to null-checking. But in Java's instances, reported in \autoref{tab:popularConstructsPerType-java}, 82.03\% of preconditions and 71.12\% of postconditions are associated with null-checking.  In this number, we are not considering potential \textit{CREIllegalArgumentException} and \textit{CREIllegalStateException} that could be associated with null-checking since this would require analyzing the condition present at the \emph{if-statement}. This confirms a lack of expressiveness in the contracts specified by Java practitioners, with most being associated with null-checking, which aligns with previous studies \citep{kn:schiller2014, kn:Estler2014}.

This contrast in null-checking contracts between Java and Kotlin is easily explained by the languages' different takes on nullability. In Kotlin, contrary to Java, regular types are non-nullable by default; therefore, in most cases, practitioners do not have the need for constructs like \emph{AndroidXNonNull} or \emph{JSR305NonNull}. On the other hand, it is interesting to observe that relaxing this constraint to allow nullable types is not a common practice since we found no meaningful use of constraints like \emph{AndroidXNullable} and similar in Kotlin.

\begin{tcolorbox}[
    enhanced jigsaw,
    sharp corners,
    boxrule=0.5pt, 
    colback=blue!5!white,   
    boxrule=0pt, 
    frame empty
 ]
\textbf{Finding 5:} In Java applications' last versions, at least 77.72\% of preconditions, 65.63\% of postconditions, and 61.24\% of class invariants are related to null-checking. In the case of Kotlin, we found only about 3.12\% of preconditions, 6.00\% of postconditions, and 0.57\% of class invariants to be performing null-checking.
\end{tcolorbox}

\begin{table}[]
\footnotesize
\renewcommand\arraystretch{1.1}
\centering
\caption{The top 10 most frequent constructs per type in the last versions of Java applications.}
\begin{tabular}[t]{cc}
    \hline
    \hline
    Pre-conditions & Post-conditions \\
    \midrule
	AndroidXNonNull (36,031) & AndroidXNonNull (10,359) \\
    AndroidXNullable (14,983) & AndroidXNullable (5,954) \\
    CREIllegalArgumentException (7,663) & AndroidSuppressLint (3,125) \\
    CREIllegalStateException (3,232) & AndroidTargetApi (1,243) \\
    CRENullPointerException (2,230) & AndroidXRequiresApi (732) \\
    GuavaPreconditionNotNull (1,021) & AndroidXWorkerThread (551) \\
    JSR305NonNull (860) & AndroidXKeep (398) \\
    AndroidXStringRes (660) & AndroidXCallSuper (380) \\
    CREIndexOutOfBoundsException (656) & AndroidXUiThread (326) \\
    JetBrainsNotNull (612) & JSR305NonNull(322) \\
    \bottomrule
\end{tabular}
\label{tab:popularConstructsPerType-java}
\end{table}
\begin{table}[]
\footnotesize
\renewcommand\arraystretch{1.1}
\centering
\caption{The top 10 most frequent constructs per type in the last versions of Kotlin applications.}
\begin{tabular}[t]{cc}
    \hline
    \hline
    Pre-conditions & Post-conditions \\
    \midrule
    AndroidXStringRes (1,142) & AndroidSuppressLint (2,289) \\
    CREIllegalStateException (772) & AndroidXVisibleForTesting (1,663) \\
    CREIllegalArgumentException (748) & AndroidXRequiresApi (720) \\
    AndroidXColorInt (523) & AndroidXWorkerThread (638) \\
    AndroidXDrawableRes (425) & AndroidXMainThread (441) \\
    AndroidXAttrRes (255) & AndroidXCallSuper (319) \\
    AndroidXColorRes (195) & AndroidXColorInt (237) \\
    AndroidXIdRes (184) & AndroidTargetApi (205) \\
    UCREUnsupportedOperationException (116) & AndroidXUiThread (195) \\
    AndroidXFloatRange (80) & AndroidXAnyThread(184) \\
    \bottomrule
\end{tabular}
\label{tab:popularConstructsPerType-kotlin}
\end{table}


\subsection{RQ2: Evolution} 

\autoref{tab:contractsbytypeWithTwoVersions} presents the number of contracts in the first and second versions by category. In general, for most cases, the number of contracts in each category increased from the first to the last version. The only categories where the number decreased were the \emph{Apache's Commons Validate} and in \emph{JSR305 annotations package} for Java. The decrease in JSR305 usage could be explained by it currently being in a dormant status, or in other words, with no activity since 2017.

\begin{table}[]
\footnotesize
\renewcommand\arraystretch{1.1}
\centering
\caption{Contract elements by type in the first and last version.}
\begin{tabular}[t]{cccccccc}
    \hline
    \hline
    & & \multicolumn{2}{ c}{contracts (1st vers.)} & \multicolumn{2}{c}{contracts (2nd vers.)} \\ 
    \cmidrule(lr){3-4} \cmidrule(lr){5-6}
    Type & category & Java & Kotlin & Java & Kotlin \\
    \midrule
    cond. runtime exc. & CRE & 10,678 & 1,161 & 14,887 & 2,071 \\ 
    unsupp. op. exc. & CRE & 203 & 26 & 308 & 116 \\ 
    java assert & assertion & 1,308 & - & 2,217 & - \\ 
    kotlin assert & assertion & - & 1,498 & - & 2,370 \\ 
    guava precond. & API & 677 & 1 & 1,121 & 9 \\ 
    commons validate & API & 11 & 0 & 3 & 0 \\ 
    spring assert & API & 0 & 0 & 1 & 0 \\ 
    JSR303, JSR349 & annotation & 0 & 0 & 0 & 0 \\ 
    JSR305 & annotation & 2,062 & 7 & 1,133 & 13 \\ 
    findbugs & annotation & 0 & 0 & 0 & 0 \\ 
    jetbrains & annotation & 714 & 40 & 1,596 & 98 \\ 
    android & annotation & 4,990 & 2,290 & 7,013 & 3,414 \\ 
    androidx & annotation & 53,721 & 6,782 & 86,212 & 13,811 \\ 
    kotlin contracts & other & - & 0 & - & 1 \\ 
    \bottomrule
\end{tabular}
\label{tab:contractsbytypeWithTwoVersions}
\end{table}

We computed some metrics to understand how the increase in the program's size relates to the number of contracts. Those metrics, listed in \autoref{tab:evolutionMetrics}, include the average and median values for the number of methods, the number of contracts, and the ratio between both 
(for the first and second versions).
The table shows that there is an average increase of about 109.936 methods per program. This is expected since the program's size tends to increase from the first to the second version. However, a more interesting insight comes from the contracts count. Although the average number of contracts per program increased, its median value decreased. This means that the dataset includes outliers with a significant rise in contract usage that considerably affected the average value.
To confirm this data, we computed the ratio between the number of contracts and the number of methods for each version of a program. Then, we computed the difference between the second and the first version's ratio for each program. The average of these differences is -0.0057, and the median is -0.0012. Although the values are very small, we conclude that the number of methods increases significantly more than the number of contracts.

\begin{table}[]
\footnotesize
\renewcommand\arraystretch{1.1}
\centering
\caption{Average and median number of methods, contracts, and their ratio for the two versions.}
\begin{tabular}[t]{ccccc}
    \hline
    \hline
    & \multicolumn{2}{c}{1st version} & \multicolumn{2}{c}{2nd version} \\ 
    \cmidrule(lr){2-3} \cmidrule(lr){4-5}
    Metric & Median & Average & Median & Average \\
    \midrule
	methods count & 310 & 972,037 & 356 & 1,081,973 \\
    contracts count & 7 & 64,938 & 6 & 79,658 \\
    contract-to-method ratio & 0.029 & 0.061 & 0.022 & 0.055 \\
    \bottomrule
\end{tabular}
\label{tab:evolutionMetrics}
\end{table}

\begin{tcolorbox}[
    enhanced jigsaw,
    sharp corners,
    boxrule=0.5pt, 
    colback=blue!5!white,   
    boxrule=0pt, 
    frame empty
 ]
\textbf{Finding 6:} Applications that use contracts continue to use them in later versions. Moreover, the total and average numbers of contracts increase, but its median decreases by a small factor. Also, the number of methods increases at a higher rate than the number of contracts.
\end{tcolorbox}

Similarly to our study, 
Dietrich et al.~\cite{kn:dietrich17} also found that the median value of the ratio does not change much. Still,
while we reported a decline between
both versions (0.029 to 0.022), they found a rise (0.021 to 0.023). This means that although both studies show general stability related to contracts usage, contrary to their study, we were not able to find a positive correlation between the increase in the number of methods and in the number of contracts.

\subsection{RQ3: Safety}
To address whether practitioners tend to misuse contracts in either program evolution or inheritance contexts, we build \textit{diff records} to be classified according to \emph{evolution patterns}. Some of these \emph{evolution patterns} are associated with a potential risk that may lead to client breaks, namely when \emph{preconditions are strengthened} or \emph{postconditions are weakened}. This process was described in more detail in \Cref{sec:evolution-study,sec:lsp-study}.
It is important to note that the analysis tool cannot precisely capture all contract changes due to the variety of constructs we are analyzing and the complexity of their semantics. This can potentially lead to under-reporting. 
Even so, 
\Cref{tab:evolution-evolution-inheritance}
still provides valuable insights into the safety of contract usage and evolution.
%
The table shows the frequency of each \emph{evolution pattern} in the context of \emph{program evolution} (third column). At first glance, we may be rushed to conclude that specifications are generally stable since the most frequent pattern is when a contract remains \emph{unchanged} from the first to the second version. Unfortunately, this is not true since the occurrences of contract changes make up more than 50\% of the patterns found. Still, overall, most of the changes are non-critical ones\,---\,including \emph{minor changes}, \emph{preconditions weakening}, and \emph{postconditions strengthening}\,---\,which is a positive finding. Less optimistic is that the second most common pattern is the case of \emph{preconditions strengthening}, one of the two cases that potentially offers risk. In summary, although many contracts remain unchanged and most changes are not critical, we still found many occurrences that can lead to potential breaks.

\begin{tcolorbox}[
    enhanced jigsaw,
    sharp corners,
    boxrule=0.5pt, 
    colback=blue!5!white,   
    boxrule=0pt, 
    frame empty
 ]
\textbf{Finding 7:} There are instances of unsafe contract changes while the program evolves, particularly cases of preconditions strengthening.
\end{tcolorbox}

Finally, \Cref{tab:evolution-evolution-inheritance} also
presents the results found for \emph{evolution patterns} in the context of \emph{inheritance} (fourth column). We observe that the \emph{preconditions strengthening} pattern makes up almost 50\% of classified instances. We also note that from the classified instances, most parts are related to contract changes which means a lack of stability in specifications. Both in the \emph{evolution} and the \emph{inheritance} study, we found low occurrences of \emph{postconditions weakening} when compared to the other classifications. Also, compared to the reports from Dietrich et al.'s study~\cite{kn:dietrich17}, our results indicate a greater ratio of \emph{preconditions strengthening} per preconditions found.

\begin{tcolorbox}[
    enhanced jigsaw,
    sharp corners,
    boxrule=0.5pt, 
    colback=blue!5!white,   
    boxrule=0pt, 
    frame empty
 ]
\textbf{Finding 8:} There are instances of unsafe contract changes in an overriding context that violate the Liskov Substitution Principle, particularly cases of preconditions strengthening.
\end{tcolorbox}


\begin{table}[]
\footnotesize
\renewcommand\arraystretch{1.1}
\centering
\caption{Contract evolution in the context of program evolution and inheritance.}
\begin{tabular}[t]{cccc}
    \hline
    \hline
    Contract Evolution & Critical & Evolution (\#) & Inheritance (\#) \\ \hline
    \midrule
    unchanged & no & 32,070 & 158 \\
    minor change & no & 199 & 1 \\
    pre-conditions weakened & no & 13,870 & 3 \\
    post-conditions strengthened & no & 9,906 & 71 \\ \hline
    pre-conditions strengthened & yes & 20,870 & 232 \\
    post-conditions weakened & yes & 5,461 & 0 \\ 
    \hline
    unclassified & ? & 8,307 & 145 \\ 
    \bottomrule
\end{tabular}
\label{tab:evolution-evolution-inheritance}
\end{table}

\section{Discussion}
In this section, we answer the research questions listed in \Cref{sec:research-questions}, we discuss the practical implications of our findings, and we outline potential threats to the validity of our work.

\subsection{Answers to Research Questions}
Given the findings reported in the previous section, we answer the
research questions posed in \Cref{sec:research-questions} as follows:

\smallskip

\noindent\noindent
\textbf{RQ1 [Contract Usage]} {\sl How and to what extent are contracts used in Android applications?} 
Contracts are concentrated
in a small number of applications. Still, when applications use contracts, annotation-based approaches are the most frequent,
with the \emph{androidx.annotation} package being the most popular. 
The use of APIs to specify contracts is rare.
This study found 
differences
between the two languages. While in Java, 63.73\% of the classified instances are preconditions, Kotlin programs display a more equally distributed selection with 38.82\% postconditions at the top. We also found that more than 50\% of the classified contracts in Java are related to null-checking, 
while in Kotlin that number is smaller than 10\%.

\smallskip

\noindent\noindent
\textbf{RQ2 [Evolution]} {\sl How does contract usage evolve in an application?}
Applications that use contracts continue to use them in later versions.
When comparing the number of contracts in both versions of the applications, on average, the number of contracts increases. Still, this is caused by some outliers that increase its usage intensively, driving up the average. In fact, the median value decreases. 
Furthermore, the contract-to-method ratio decreases between versions --- a median decrease of -0.0057 and an average decrease of -0.0012. Although by a residual factor, we observed that the number of contracts declines as the program grew.

\smallskip

\noindent\noindent
\textbf{RQ3 [Safety]} {\sl Are contracts used safely in the context of program evolution and inheritance?} 
  We found that contract changes are frequent, meaning a lack of specifications' stability. From those changes, preconditions strengthening is the most classified pattern. These results clearly show a potentially unsafe use of contracts that may lead to client breaks. 

\subsection{Practical Implications \& Recommendations}
From the findings presented in this work, we are able to derive practical implications and recommendations.

%
\smallskip

\noindent\noindent
\textbf{Recommendation 1:} Due to the visible fragmentation of technologies and approaches to specifying contracts, both Java and Kotlin standard libraries should be equipped with specialized constructs to specify contracts and with proper official documentation. 

\smallskip

\noindent\noindent
\textbf{Recommendation 2:} It would be desirable to have libraries that standardize contract specifications in  Java and Kotlin. Our results suggest that these libraries should be built around annotation-based contracts, given its popularity among practitioners. An annotation-based approach, where specifications are added to the program as metadata, is similar to Eiffel's approach, where the assertions do not obfuscate the method's implementation. This recommendation also applies to tool builders: given that the current use of APIs in Android development appears to be relatively low, analysis tools for Android that leverage contracts should prioritize annotations. 

\smallskip

\noindent\noindent
\textbf{Recommendation 3:} New tools to aid practitioners writing contracts would be very valuable. For example, the integration into IDEs of contract suggestion features supported by tools for invariant inference, such as Daikon~\cite{ernst2007daikon}, could help increase practitioners' use of contracts. Another contribution could be IDE and continuous integration plugins to detect contract violations in the context of program evolution and inheritance.

\smallskip

\noindent\noindent
\textbf{Recommendation 4:} Our findings show that Kotlin's default non-nullable types reduce the need to explicitly write some contracts, highlighting the significance of language design features that enable safety by default. 
These findings are relevant for the design of programming languages and can serve as motivation for practitioners when selecting programming languages for new projects.

\subsection{Threats to Validity}
{\bf \emph{Internal Validity.}} The accuracy of our results depends on the quality and correctness of the artifact and there may exist bugs in the code. To mitigate this, we extensively tested the tool. In addition, all code and datasets used are publicly available for other researchers and potential users to check the validity of the results.\\
%
{\bf \emph{External Validity.}} The projects that we selected might not be an accurate representation of other, more popular, Android app stores. We mitigated this by using F-Droid, a collection of open-source applications commonly used in other research studies. We also mitigated this risk by analysing \emph{all} the projects that satisfy the inclusion criteria, leading to a substantial dataset (51 MLoC) with applications of different types.\\
%
%
{\bf\emph{Conclusion Validity.}} We might have missed language constructs that could be used to specify contracts. To mitigate this, we followed an established taxonomy~\cite{kn:dietrich17} that we adapted and extended by systematically searching forums and the official Android documentation. Also, all our code is 
easily open to extension. Another risk comes from the fact that our dataset is imbalanced (with more Java than Kotlin applications). We mitigate this by explicitly discussing this imbalance when presenting results that might be affected by it.


\section{Conclusions} 
Additional empirical evidence about what types of constructs and language features are used by practitioners to represent contracts can help the software engineering community create or improve existing libraries and tools to increase DbC adoption. This knowledge also serves practitioners to understand DbC's current practices better, helping them discover and decide between different implementation approaches of this technique for their projects. Other researchers can also use this data to draw additional studies and to foster further discussion along with the increasing interest in these empirical studies about language features.

Future work includes studies with practitioners to understand the challenges faced when specifying contracts, the use of annotations to improve Android analysis tools \cite{ribeiro2021ecoandroid,pereira2022extending}, and the development of tools that can help increase the adoption of DbC \cite{silva2024leveraging}.


%
\bibliographystyle{ACM-Reference-Format}
\bibliography{refs}


\begin{thebibliography}{36}


\ifx \showCODEN    \undefined \def \showCODEN     #1{\unskip}     \fi
\ifx \showDOI      \undefined \def \showDOI       #1{#1}\fi
\ifx \showISBNx    \undefined \def \showISBNx     #1{\unskip}     \fi
\ifx \showISBNxiii \undefined \def \showISBNxiii  #1{\unskip}     \fi
\ifx \showISSN     \undefined \def \showISSN      #1{\unskip}     \fi
\ifx \showLCCN     \undefined \def \showLCCN      #1{\unskip}     \fi
\ifx \shownote     \undefined \def \shownote      #1{#1}          \fi
\ifx \showarticletitle \undefined \def \showarticletitle #1{#1}   \fi
\ifx \showURL      \undefined \def \showURL       {\relax}        \fi
\providecommand\bibfield[2]{#2}
\providecommand\bibinfo[2]{#2}
\providecommand\natexlab[1]{#1}
\providecommand\showeprint[2][]{arXiv:#2}

\bibitem[A.Feldman et~al\mbox{.}(2006)]%
        {kn:feldman2006}
\bibfield{author}{\bibinfo{person}{Y. A.Feldman}, \bibinfo{person}{O.
  Barzilay}, {and} \bibinfo{person}{S. Tyszberowicz}.}
  \bibinfo{year}{2006}\natexlab{}.
\newblock \showarticletitle{Jose: aspects for design by contract}. In
  \bibinfo{booktitle}{\emph{Fourth IEEE International Conference on Software
  Engineering and Formal Methods}}. \bibinfo{address}{Los Alamitos, CA, USA}.
\newblock


\bibitem[Algarni and Magel(2018)]%
        {kn:algarni2018}
\bibfield{author}{\bibinfo{person}{A. Algarni} {and} \bibinfo{person}{K.
  Magel}.} \bibinfo{year}{2018}\natexlab{}.
\newblock \showarticletitle{Toward Design-by-Contract Based Generative Tool for
  Object-Oriented System}. In \bibinfo{booktitle}{\emph{2018 IEEE 9th
  International Conference on Software Engineering and Service Science
  (ICSESS). Proceedings}}. \bibinfo{address}{Piscataway, NJ, USA},
  \bibinfo{pages}{168 -- 73}.
\newblock


\bibitem[Aniche(2022)]%
        {kn:Aniche2022}
\bibfield{author}{\bibinfo{person}{M. Aniche}.}
  \bibinfo{year}{2022}\natexlab{}.
\newblock \bibinfo{booktitle}{\emph{Effective Software Testing. A Developer's
  Guide.}}
\newblock \bibinfo{publisher}{Manning}, \bibinfo{address}{Shelter Islands}.
\newblock
\showISBNx{9781633439931}


\bibitem[Backes et~al\mbox{.}(2016)]%
        {kn:backes2016}
\bibfield{author}{\bibinfo{person}{M. Backes}, \bibinfo{person}{S. Bugiel},
  {and} \bibinfo{person}{E. Derr}.} \bibinfo{year}{2016}\natexlab{}.
\newblock \showarticletitle{Reliable Third-Party Library Detection in Android
  and Its Security Applications}. In \bibinfo{booktitle}{\emph{Proceedings of
  the 2016 ACM SIGSAC Conference on Computer and Communications Security}}
  \emph{(\bibinfo{series}{CCS '16})}. \bibinfo{publisher}{Association for
  Computing Machinery}, \bibinfo{pages}{356–367}.
\newblock
\showISBNx{9781450341394}


\bibitem[Bloch(2008)]%
        {bloch2008effective}
\bibfield{author}{\bibinfo{person}{Joshua Bloch}.}
  \bibinfo{year}{2008}\natexlab{}.
\newblock \bibinfo{booktitle}{\emph{Effective java} (\bibinfo{edition}{2nd}
  ed.)}.
\newblock \bibinfo{publisher}{Addison-Wesley Professional}.
\newblock


\bibitem[Blom et~al\mbox{.}(2002a)]%
        {kn:Blom2002}
\bibfield{author}{\bibinfo{person}{M. Blom}, \bibinfo{person}{E.J. Nordby},
  {and} \bibinfo{person}{A. Brunstrom}.} \bibinfo{year}{2002}\natexlab{a}.
\newblock \showarticletitle{An experimental evaluation of programming by
  contract}. In \bibinfo{booktitle}{\emph{Proceedings Ninth Annual IEEE
  International Conference and Workshop on the Engineering of Computer-Based
  Systems}}. \bibinfo{pages}{118--127}.
\newblock


\bibitem[Blom et~al\mbox{.}(2002b)]%
        {kn:Blom20022}
\bibfield{author}{\bibinfo{person}{M. Blom}, \bibinfo{person}{E.~J. Nordby},
  {and} \bibinfo{person}{A. Brunstrom}.} \bibinfo{year}{2002}\natexlab{b}.
\newblock \showarticletitle{On the relation between design contracts and
  errors: a software development strategy}. In
  \bibinfo{booktitle}{\emph{Proceedings Ninth Annual IEEE International
  Conference and Workshop on the Engineering of Computer-Based Systems}}.
  \bibinfo{pages}{110--117}.
\newblock


\bibitem[Casalnuovo et~al\mbox{.}(2015)]%
        {kn:casalnuovo2015}
\bibfield{author}{\bibinfo{person}{C. Casalnuovo}, \bibinfo{person}{P.
  Devanbu}, \bibinfo{person}{A. Oliveira}, \bibinfo{person}{V. Filkov}, {and}
  \bibinfo{person}{B. Ray}.} \bibinfo{year}{2015}\natexlab{}.
\newblock \showarticletitle{Assert Use in GitHub Projects}. In
  \bibinfo{booktitle}{\emph{2015 IEEE/ACM 37th IEEE International Conference on
  Software Engineering (ICSE). Proceedings}}, Vol.~\bibinfo{volume}{1}.
  \bibinfo{address}{Los Alamitos, CA, USA}, \bibinfo{pages}{755 -- 66}.
\newblock


\bibitem[Chalin(2006)]%
        {kn:chalin2006}
\bibfield{author}{\bibinfo{person}{P. Chalin}.}
  \bibinfo{year}{2006}\natexlab{}.
\newblock \bibinfo{booktitle}{\emph{Are practitioners writing contracts?}}
\newblock \bibinfo{publisher}{Springer}, \bibinfo{address}{Berlin, Germany},
  \bibinfo{pages}{100 -- 113}.
\newblock
\showISBNx{978-3-540-48265-9}


\bibitem[Chen et~al\mbox{.}(2019)]%
        {chen2019storydroid}
\bibfield{author}{\bibinfo{person}{Sen Chen}, \bibinfo{person}{Lingling Fan},
  \bibinfo{person}{Chunyang Chen}, \bibinfo{person}{Ting Su},
  \bibinfo{person}{Wenhe Li}, \bibinfo{person}{Yang Liu}, {and}
  \bibinfo{person}{Lihua Xu}.} \bibinfo{year}{2019}\natexlab{}.
\newblock \showarticletitle{Storydroid: Automated generation of storyboard for
  Android apps}. In \bibinfo{booktitle}{\emph{2019 IEEE/ACM 41st International
  Conference on Software Engineering (ICSE)}}. IEEE, \bibinfo{pages}{596--607}.
\newblock


\bibitem[Counsell et~al\mbox{.}(2017)]%
        {kn:counsell2017}
\bibfield{author}{\bibinfo{person}{S. Counsell}, \bibinfo{person}{T. Hall},
  \bibinfo{person}{T. Shippey}, \bibinfo{person}{D. Bowes}, \bibinfo{person}{A.
  Tahir}, {and} \bibinfo{person}{S. MacDonell}.}
  \bibinfo{year}{2017}\natexlab{}.
\newblock \showarticletitle{Assert Use and Defectiveness in Industrial Code}.
  In \bibinfo{booktitle}{\emph{Proceedings of the IEEE International Symposium
  on Software Reliability Engineering Workshops}}. \bibinfo{pages}{20--23}.
\newblock


\bibitem[Dietrich et~al\mbox{.}(2017)]%
        {kn:dietrich17}
\bibfield{author}{\bibinfo{person}{J. Dietrich}, \bibinfo{person}{D.~J.
  Pearce}, \bibinfo{person}{K. Jezek}, {and} \bibinfo{person}{P. Brada}.}
  \bibinfo{year}{2017}\natexlab{}.
\newblock \showarticletitle{Contracts in the Wild: A Study of Java Programs}.
  In \bibinfo{booktitle}{\emph{31st European Conference on Object-Oriented
  Programming (ECOOP 2017)}} \emph{(\bibinfo{series}{Leibniz International
  Proceedings in Informatics (LIPIcs)}, Vol.~\bibinfo{volume}{74})}.
  \bibinfo{publisher}{Schloss Dagstuhl--Leibniz-Zentrum fuer Informatik},
  \bibinfo{address}{Dagstuhl, Germany}, \bibinfo{pages}{9:1--9:29}.
\newblock
\showISBNx{978-3-95977-035-4}
\showISSN{1868-8969}


\bibitem[Ernst et~al\mbox{.}(2007)]%
        {ernst2007daikon}
\bibfield{author}{\bibinfo{person}{Michael~D Ernst}, \bibinfo{person}{Jeff~H
  Perkins}, \bibinfo{person}{Philip~J Guo}, \bibinfo{person}{Stephen McCamant},
  \bibinfo{person}{Carlos Pacheco}, \bibinfo{person}{Matthew~S Tschantz}, {and}
  \bibinfo{person}{Chen Xiao}.} \bibinfo{year}{2007}\natexlab{}.
\newblock \showarticletitle{The {Daikon} system for dynamic detection of likely
  invariants}.
\newblock \bibinfo{journal}{\emph{Sci. Comput. Program.}} \bibinfo{volume}{69},
  \bibinfo{number}{1-3} (\bibinfo{year}{2007}), \bibinfo{pages}{35--45}.
\newblock


\bibitem[Estler et~al\mbox{.}(2014)]%
        {kn:Estler2014}
\bibfield{author}{\bibinfo{person}{H.-C. Estler}, \bibinfo{person}{C.~A.
  Furia}, \bibinfo{person}{M. Nordio}, \bibinfo{person}{M. Piccioni}, {and}
  \bibinfo{person}{B. Meyer}.} \bibinfo{year}{2014}\natexlab{}.
\newblock \showarticletitle{Contracts in Practice}. In
  \bibinfo{booktitle}{\emph{FM 2014: Formal Methods. 19th International
  Symposium. Proceedings: LNCS 8442}}. \bibinfo{address}{Cham, Switzerland},
  \bibinfo{pages}{230 -- 46}.
\newblock


\bibitem[Fairbanks(2019)]%
        {kn:Fairbanks19}
\bibfield{author}{\bibinfo{person}{G. Fairbanks}.}
  \bibinfo{year}{2019}\natexlab{}.
\newblock \showarticletitle{Better code reviews with design by contract}.
\newblock \bibinfo{journal}{\emph{IEEE Software}} \bibinfo{volume}{36},
  \bibinfo{number}{6} (\bibinfo{year}{2019}), \bibinfo{pages}{53 -- 6}.
\newblock


\bibitem[Grazia and Pradel(2022)]%
        {kn:grazia2022}
\bibfield{author}{\bibinfo{person}{L.~Di Grazia} {and} \bibinfo{person}{M.
  Pradel}.} \bibinfo{year}{2022}\natexlab{}.
\newblock \showarticletitle{The Evolution of Type Annotations in Python: An
  Empirical Study}. In \bibinfo{booktitle}{\emph{Proceedings of the 30th ACM
  Joint European Software Engineering Conference and Symposium on the
  Foundations of Software Engineering}} (Singapore, Singapore).
  \bibinfo{publisher}{Association for Computing Machinery},
  \bibinfo{address}{New York, NY, USA}, \bibinfo{pages}{209–220}.
\newblock
\showISBNx{9781450394130}


\bibitem[Hollunder et~al\mbox{.}(2012)]%
        {kn:Hollunder2012}
\bibfield{author}{\bibinfo{person}{B. Hollunder}, \bibinfo{person}{M.
  Herrmann}, {and} \bibinfo{person}{A. H{\"u}lzenbecher}.}
  \bibinfo{year}{2012}\natexlab{}.
\newblock \showarticletitle{Design by Contract for Web Services: Architecture,
  Guidelines, and Mappings}. In \bibinfo{booktitle}{\emph{International Journal
  on Advances in Software}}, Vol.~\bibinfo{volume}{5}.
\newblock


\bibitem[Kochhar and Lo(2017)]%
        {kn:kochar2017}
\bibfield{author}{\bibinfo{person}{P. Kochhar} {and} \bibinfo{person}{D. Lo}.}
  \bibinfo{year}{2017}\natexlab{}.
\newblock \showarticletitle{Revisiting Assert Use in GitHub Projects}. In
  \bibinfo{booktitle}{\emph{Proceedings of the 21st International Conference on
  Evaluation and Assessment in Software Engineering}}.
  \bibinfo{pages}{298--307}.
\newblock


\bibitem[Kudrjavets et~al\mbox{.}(2006)]%
        {kn:kudrjavets2006}
\bibfield{author}{\bibinfo{person}{Gunnar Kudrjavets},
  \bibinfo{person}{Nachiappan Nagappan}, {and} \bibinfo{person}{Thomas Ball}.}
  \bibinfo{year}{2006}\natexlab{}.
\newblock \showarticletitle{Assessing the Relationship between Software
  Assertions and Faults: An Empirical Investigation}. In
  \bibinfo{booktitle}{\emph{2006 17th International Symposium on Software
  Reliability Engineering}}. \bibinfo{pages}{204--212}.
\newblock


\bibitem[Meyer(1988)]%
        {meyer1988programming}
\bibfield{author}{\bibinfo{person}{Bertrand Meyer}.}
  \bibinfo{year}{1988}\natexlab{}.
\newblock \showarticletitle{Programming as contracting}.
\newblock \bibinfo{journal}{\emph{Advances in Object-Oriented Software
  Engineering}} (\bibinfo{year}{1988}), \bibinfo{pages}{1--15}.
\newblock


\bibitem[Meyer(1992)]%
        {kn:Meyer1992}
\bibfield{author}{\bibinfo{person}{B. Meyer}.} \bibinfo{year}{1992}\natexlab{}.
\newblock \showarticletitle{Applying `design by contract'}.
\newblock \bibinfo{journal}{\emph{Computer}} \bibinfo{volume}{25},
  \bibinfo{number}{10} (\bibinfo{year}{1992}), \bibinfo{pages}{40 -- 51}.
\newblock
\showISSN{0018-9162}


\bibitem[Murthy(2018)]%
        {kn:murthy2019}
\bibfield{author}{\bibinfo{person}{P.~V.~R. Murthy}.}
  \bibinfo{year}{2018}\natexlab{}.
\newblock \showarticletitle{Design by contract methodology}. In
  \bibinfo{booktitle}{\emph{2018 International Conference on Advances in
  Computing, Communications and Informatics (ICACCI)}}.
  \bibinfo{address}{Piscataway, NJ, USA}, \bibinfo{pages}{482 -- 8}.
\newblock


\bibitem[Naumchev(2019)]%
        {kn:naumchev2019}
\bibfield{author}{\bibinfo{person}{A. Naumchev}.}
  \bibinfo{year}{2019}\natexlab{}.
\newblock \showarticletitle{Seamless Object-Oriented Requirements}. In
  \bibinfo{booktitle}{\emph{2019 International Multi-Conference on Engineering,
  Computer and Information Sciences (SIBIRCON). Proceedings}}.
  \bibinfo{address}{Piscataway, NJ, USA}.
\newblock


\bibitem[Pereira et~al\mbox{.}(2022)]%
        {pereira2022extending}
\bibfield{author}{\bibinfo{person}{Ricardo~B Pereira},
  \bibinfo{person}{Jo{\~a}o~F. Ferreira}, \bibinfo{person}{Alexandra Mendes},
  {and} \bibinfo{person}{Rui Abreu}.} \bibinfo{year}{2022}\natexlab{}.
\newblock \showarticletitle{Extending {E}coandroid with automated detection of
  resource leaks}. In \bibinfo{booktitle}{\emph{Proceedings of the 9th IEEE/ACM
  International Conference on Mobile Software Engineering and Systems}}.
  \bibinfo{pages}{17--27}.
\newblock


\bibitem[Ribeiro et~al\mbox{.}(2021)]%
        {ribeiro2021ecoandroid}
\bibfield{author}{\bibinfo{person}{Ana Ribeiro}, \bibinfo{person}{Jo{\~a}o~F.
  Ferreira}, {and} \bibinfo{person}{Alexandra Mendes}.}
  \bibinfo{year}{2021}\natexlab{}.
\newblock \showarticletitle{Ecoandroid: An {A}ndroid studio plugin for
  developing energy-efficient {J}ava mobile applications}. In
  \bibinfo{booktitle}{\emph{2021 IEEE 21st International Conference on Software
  Quality, Reliability and Security (QRS)}}. IEEE, \bibinfo{pages}{62--69}.
\newblock


\bibitem[Schiller et~al\mbox{.}(2014)]%
        {kn:schiller2014}
\bibfield{author}{\bibinfo{person}{T.~W. Schiller}, \bibinfo{person}{K.
  Donohue}, \bibinfo{person}{F. Coward}, {and} \bibinfo{person}{M.~D. Ernst}.}
  \bibinfo{year}{2014}\natexlab{}.
\newblock \showarticletitle{Case Studies and Tools for Contract
  Specifications}. In \bibinfo{booktitle}{\emph{Proceedings of the 36th
  International Conference on Software Engineering}} (Hyderabad, India)
  \emph{(\bibinfo{series}{ICSE 2014})}. \bibinfo{publisher}{Association for
  Computing Machinery}, \bibinfo{address}{New York, NY, USA},
  \bibinfo{pages}{596–607}.
\newblock
\showISBNx{9781450327565}


\bibitem[Silva et~al\mbox{.}(2020)]%
        {kn:silva2020}
\bibfield{author}{\bibinfo{person}{C. Silva}, \bibinfo{person}{S. Guerin},
  \bibinfo{person}{R. Mazo}, {and} \bibinfo{person}{J. Champeau}.}
  \bibinfo{year}{2020}\natexlab{}.
\newblock \showarticletitle{Contract-based design patterns: a design by
  contract approach to specify security patterns}. In
  \bibinfo{booktitle}{\emph{ARES 2020: Proceedings of the 15th International
  Conference on Availability, Reliability and Security}}. \bibinfo{address}{New
  York, NY, USA}.
\newblock


\bibitem[Stats(2023)]%
        {kn:OSUsage2023}
\bibfield{author}{\bibinfo{person}{StatCounter~Global Stats}.}
  \bibinfo{year}{2023}\natexlab{}.
\newblock \bibinfo{title}{Operating System Market Share Worldwide}.
\newblock
\newblock
\urldef\tempurl%
\url{https://gs.statcounter.com/os-market-share#monthly-202208-202209-bar}
\showURL{%
\tempurl}
\newblock
\shownote{[Online; accessed 3-February-2023]}.


\bibitem[Tantivongsathaporn and Stearns(2006)]%
        {kn:Tantivongsathaporn2006}
\bibfield{author}{\bibinfo{person}{J. Tantivongsathaporn} {and}
  \bibinfo{person}{D. Stearns}.} \bibinfo{year}{2006}\natexlab{}.
\newblock \showarticletitle{An experience with design by contract}. In
  \bibinfo{booktitle}{\emph{2006 13th Asia Pacific Software Engineering
  Conference (APSEC'06)}}. \bibinfo{address}{Piscataway, NJ, USA},
  \bibinfo{pages}{327 -- 33}.
\newblock


\bibitem[Tao and Edmunds(2018)]%
        {kn:TaoEdmunds18}
\bibfield{author}{\bibinfo{person}{K. Tao} {and} \bibinfo{person}{P. Edmunds}.}
  \bibinfo{year}{2018}\natexlab{}.
\newblock \showarticletitle{Mobile APPs and Global Markets}.
\newblock \bibinfo{journal}{\emph{Theoretical Economics Letters}}
  \bibinfo{volume}{08} (\bibinfo{date}{01} \bibinfo{year}{2018}),
  \bibinfo{pages}{1510--1524}.
\newblock


\bibitem[Wang et~al\mbox{.}(2020)]%
        {kn:wang2020}
\bibfield{author}{\bibinfo{person}{Y. Wang}, \bibinfo{person}{B. Chen},
  \bibinfo{person}{K. Huang}, \bibinfo{person}{B. Shi}, \bibinfo{person}{C.
  Xu}, \bibinfo{person}{X. Peng}, \bibinfo{person}{Y. Wu}, {and}
  \bibinfo{person}{Y. Liu}.} \bibinfo{year}{2020}\natexlab{}.
\newblock \showarticletitle{An Empirical Study of Usages, Updates and Risks of
  Third-Party Libraries in Java Projects}. In \bibinfo{booktitle}{\emph{2020
  IEEE International Conference on Software Maintenance and Evolution
  (ICSME)}}. \bibinfo{pages}{35--45}.
\newblock


\bibitem[Wei et~al\mbox{.}(2011)]%
        {kn:furiaWeiKazmin2011}
\bibfield{author}{\bibinfo{person}{Y. Wei}, \bibinfo{person}{C.A. Furia},
  \bibinfo{person}{N. Kazmin}, {and} \bibinfo{person}{B. Meyer}.}
  \bibinfo{year}{2011}\natexlab{}.
\newblock \showarticletitle{Inferring better contracts}. In
  \bibinfo{booktitle}{\emph{2011 33rd International Conference on Software
  Engineering (ICSE 2011)}}. \bibinfo{address}{Piscataway, NJ, USA},
  \bibinfo{pages}{191 -- 200}.
\newblock


\bibitem[Yu et~al\mbox{.}(2021)]%
        {kn:yu2021}
\bibfield{author}{\bibinfo{person}{Z. Yu}, \bibinfo{person}{C. Bai},
  \bibinfo{person}{L. Seinturier}, {and} \bibinfo{person}{M. Monperrus}.}
  \bibinfo{year}{2021}\natexlab{}.
\newblock \showarticletitle{Characterizing the Usage, Evolution and Impact of
  Java Annotations in Practice}.
\newblock \bibinfo{journal}{\emph{IEEE Transactions on Software Engineering}}
  \bibinfo{volume}{47}, \bibinfo{number}{5} (\bibinfo{year}{2021}),
  \bibinfo{pages}{969--986}.
\newblock


\bibitem[Zeng et~al\mbox{.}(2019)]%
        {zeng2019studying}
\bibfield{author}{\bibinfo{person}{Yi Zeng}, \bibinfo{person}{Jinfu Chen},
  \bibinfo{person}{Weiyi Shang}, {and} \bibinfo{person}{Tse-Hsun Chen}.}
  \bibinfo{year}{2019}\natexlab{}.
\newblock \showarticletitle{Studying the characteristics of logging practices
  in mobile apps: a case study on f-droid}.
\newblock \bibinfo{journal}{\emph{Empirical Software Engineering}}
  \bibinfo{volume}{24} (\bibinfo{year}{2019}), \bibinfo{pages}{3394--3434}.
\newblock


\bibitem[Zhou et~al\mbox{.}(2017)]%
        {kn:Zhou2017}
\bibfield{author}{\bibinfo{person}{Y. Zhou}, \bibinfo{person}{P. Pelliccione},
  \bibinfo{person}{J. Haraldsson}, {and} \bibinfo{person}{M. Islam}.}
  \bibinfo{year}{2017}\natexlab{}.
\newblock \showarticletitle{Improving Robustness of AUTOSAR Software Components
  with Design by Contract: A Study Within Volvo AB}. In
  \bibinfo{booktitle}{\emph{Software Engineering for Resilient Systems. 9th
  International Workshop, SERENE 2017. Proceedings: LNCS 10479}}.
  \bibinfo{address}{Cham, Switzerland}, \bibinfo{pages}{151 -- 68}.
\newblock


\bibitem[Álvaro Silva et~al\mbox{.}(2024)]%
        {silva2024leveraging}
\bibfield{author}{\bibinfo{person}{Álvaro Silva}, \bibinfo{person}{Alexandra
  Mendes}, {and} \bibinfo{person}{João~F. Ferreira}.}
  \bibinfo{year}{2024}\natexlab{}.
\newblock \bibinfo{title}{Leveraging Large Language Models to Boost Dafny's
  Developers Productivity}.
\newblock
\newblock
\showeprint[arxiv]{2401.00963}~[cs.SE]


\end{thebibliography}

\appendix
\clearpage
\onecolumn
\section*{Supplementary Material}
The following appendices contain additional material that complements the main body of the paper. 

\section{Information about Available Data / Artifacts}
All the code and datasets are publicly available at \url{https://figshare.com/s/d6eb7e5522bb535dc81a}



\medskip

\noindent Alternatively, the following links can also be used:
\begin{itemize}
\item The code is available at: \url{https://anonymous.4open.science/r/contracts-android-3E30}
\item The dataset is available at: \url{https://drive.google.com/file/d/1X8Qy3yamzjIZyc_h5AtNW84-91qFTiV4/view?usp=share_link}
\end{itemize}

\section{Dataset: GitHub Statistics}
\begin{figure}[!ht]
    \centering
    \begin{subfigure}{0.4\columnwidth}
        \centering
        \includegraphics[width=\columnwidth]{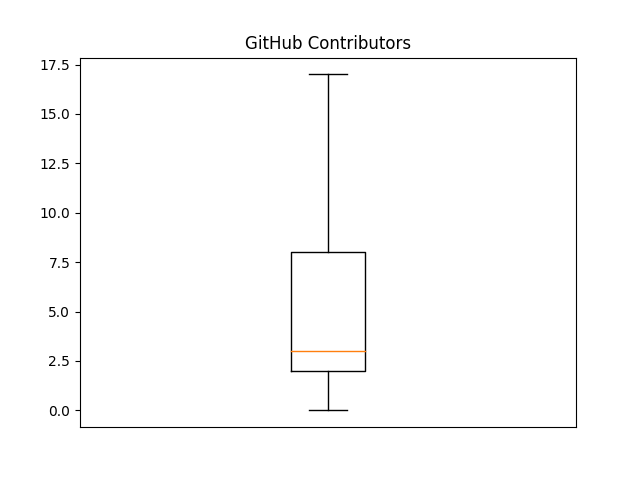}
        \caption{Number of contributors.}
        \label{fig:4-dataset-contributors}
    \end{subfigure}%
    \begin{subfigure}{0.4\columnwidth}
        \centering
        \includegraphics[width=\columnwidth]{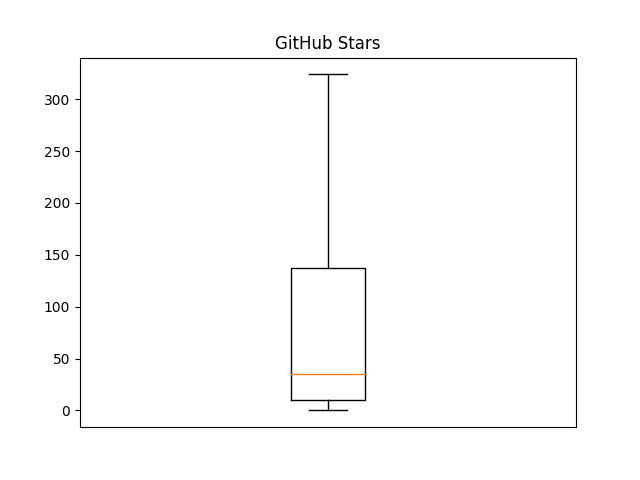}
        \caption{Number of stars.}
        \label{fig:4-dataset-stars}
    \end{subfigure}
    \begin{subfigure}{0.4\columnwidth}
        \centering
        \includegraphics[width=\columnwidth]{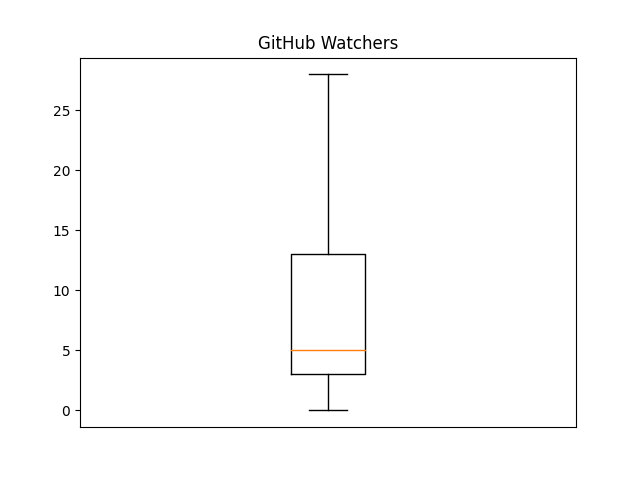}
        \caption{Number of watchers.}
        \label{fig:4-dataset-watchers}
    \end{subfigure}
    \begin{subfigure}{0.4\columnwidth}
        \centering
        \includegraphics[width=\columnwidth]{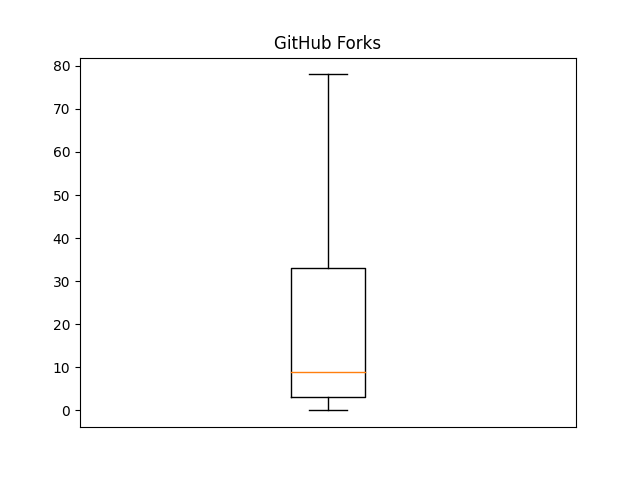}
        \caption{Number of forks.}
        \label{fig:4-dataset-forks}
    \end{subfigure}
    \caption{The distribution of GitHub repositories-related metrics for the dataset's projects, without outliers.}
    \label{fig:4-github-metrics}
\end{figure}

\autoref{fig:4-github-metrics} shows the distribution of GitHub-related metrics --- including the number of \emph{contributors}, \emph{stars}, \emph{watchers}, and \emph{forks} --- for the projects that form the evaluated dataset. While the number of \emph{contributors} describes the project's team and its size, the number of \emph{stars}, \emph{watchers}, and \emph{forks} help to assess each project's popularity and relevance among other developers. For reference, the maximum outlier for each metric is 1682 for watchers, 33,689 for stars, 11,633 for forks, and 398 for contributors. Again, this diversity ensures the quality of the dataset and reduces potential bias.

\newpage
\section{Contracts in Android Applications}
\subsection{CREs}
An exception can be used to signal, at runtime, a contract violation. However, it is important to note that the exception itself does not represent a contract; it needs to be associated with a previous check --- for example, an exception thrown inside an \emph{if-else block} --- to be considered so. Java and Kotlin offer many exceptions that can be used for this purpose, such as the \emph{IllegalArgumentException}. The \emph{android.util} package offers additional exceptions that we are also interested in analyzing, such as the case of the \emph{AndroidRuntimeException}. 
We are also interested in a particular exception, the \textit{UnsupportedOperationException}, which, according to the Java documentation, is thrown to indicate that the requested operation is not supported. 
An example of this is the following method \emph{proceedWithCheckout}, which states that it can only perform its task when the \emph{shoppingCart} has at least one item:

\begin{lstlisting}[
language=Java, 
label={lst:3-java-cre},
numbers=none,
aboveskip=1em,
belowskip=1em,
frame=none,
backgroundcolor=\color{white}
]
  public void proceedWithCheckout(List<Item> shoppingCart)  {
    if (shoppingCart.isEmpty()) {
      throw new IllegalArgumentException();
    }
    ...
  }
\end{lstlisting}

\subsection{APIs}
The methods provided by the \emph{Validate} class are simply wrapping exceptions that we have already considered in the CREs. 
Still, as we can see in the following Kotlin code snippet, this API contributes to cleaner code compared to a raw CRE-based solution since we can specify the contracts in a single line and with meaningful wording. In particular, we specify a precondition stating that \emph{the items list is not empty}:
\begin{lstlisting}[
language=Java, 
label={lst:3-api-validate},
numbers=none,
aboveskip=1em,
belowskip=1em,
frame=none,
backgroundcolor=\color{white}
]   
  fun addToShoppingCart(items: List<Item>): List<Item> {
    Validate.notEmpty(items)
    shoppingCart.addAll(items)
    return shoppingCart
  }
\end{lstlisting}

\newpage

\section{Algorithm for diff records of the contracts} 
\begin{figure}[ht]
    \centering
    \includegraphics[width=0.5\columnwidth]{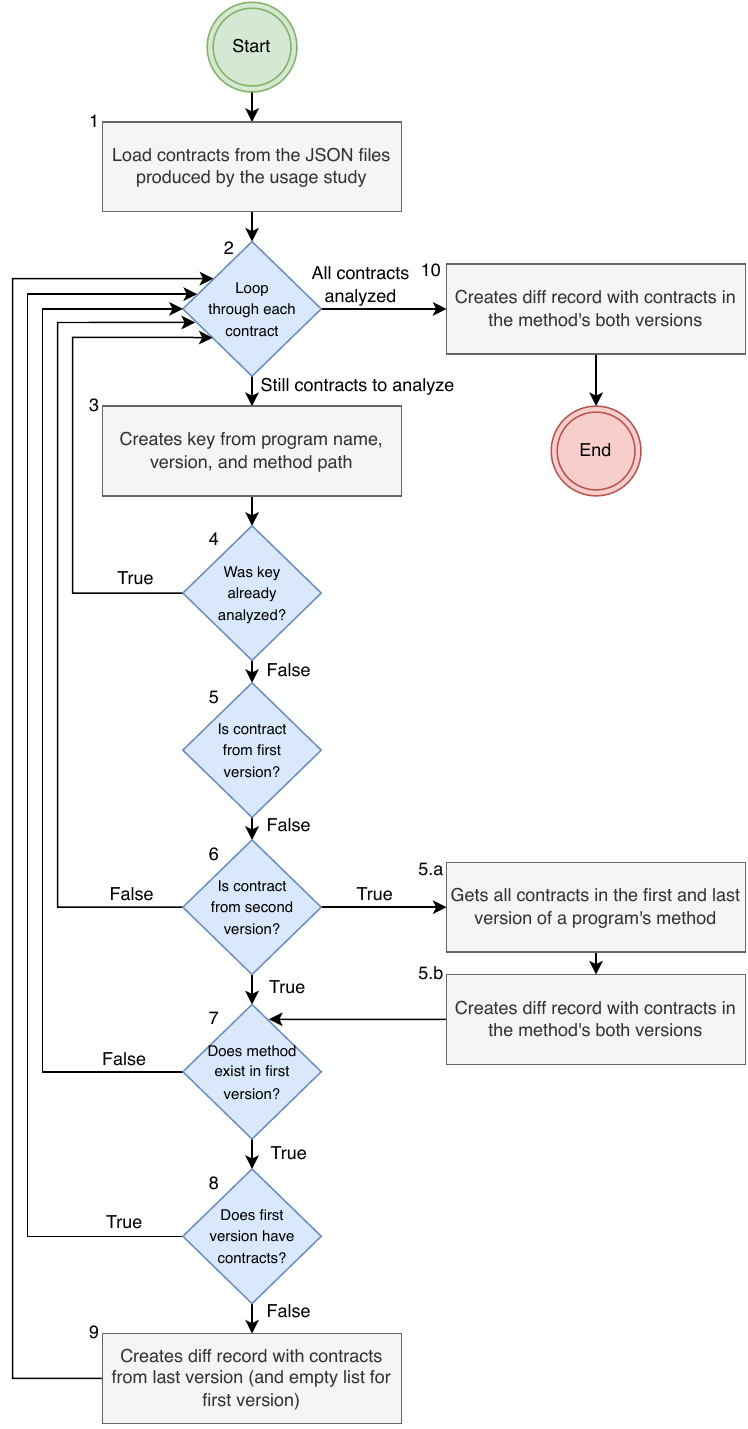}
    \caption{An overview of the algorithm to create \emph{diff records} of the contracts found in two versions of a method.}
    \label{fig:3-evolution-analyze-flowchart}
\end{figure}

The algorithm to create \emph{diff records} is illustrated in
\autoref{fig:3-evolution-analyze-flowchart}. 
The algorithm to create \emph{diff records} consists of walking through each contract identified in the \emph{usage} study (steps 1 and 2). 
We create a unique index for each contract in the loop to ensure we are not double-counting occurrences (steps 3 and 4). 
If the contract was not analyzed yet, we determine whether the contract belongs to the first version of the application (step 5). 
If this is the case, we create a \emph{diff record} by retrieving all the contracts in both versions of this contract's method (steps 5.a and 5.b). 
Otherwise, if the contract belongs to the last version of the application (step 6),
we determine whether the associated method already existed in the first version (step 7). 
If the method existed and its first version did not contain contracts, we create a diff record with only the last version's contracts (steps 8 and 9). 
If the first version contains contracts, we do not create a \emph{diff record} to not double-count contracts since they will be reported by step 5.b. 
Ultimately, the program outputs the diff records created for each program-version method (step 10).

\clearpage
\newpage
\section{Liskov Substitution Principle Study: Example}
~\autoref{lst:3-evolution-inheritance-1} shows the \emph{TagEntry} class that extends the \emph{EntryItem} class. It also overrides the \emph{setName} method inherited from its parent class. Note that while the superclass implementation contains no contract, the subclass implementation adds a CRE precondition throwing an \emph{IllegalStateException} when the \emph{id} property does not end with the \emph{name} parameter. Therefore, we are in the presence of a \emph{precondition strengthening} in the context of \emph{inheritance}, i.e., a violation of the Liskov Substitution Principle.

\bigskip

\begin{lstlisting}[
language=Java, 
caption={A Java class that overrides the setName method from its parent class.
This is an example of \emph{pre-condition strengthening} in the context of \emph{inheritance}, i.e., a violation of the Liskov Substitution Principle.}, 
captionpos=b,
label={lst:3-evolution-inheritance-1},
morekeywords={@Override},
belowskip=0pt,
aboveskip=0pt
]
public class EntryItem {
  public void setName(String name) {
    if (name != null) {
      this.name = name;
      this.normalizedName = StringNormalizer.normalizeWithResult(this.name, false);
    } else {
      this.name = "null";
      this.normalizedName = null;
    }
  }
}
    
public class TagEntry extends EntryItem {      
  public final String id;
        
  @Override
  public void setName(String name) {
    if (name != null) {
      if (!id.endsWith(name))
        throw new IllegalStateException("The display name and tag name need to be equal.");
      super.setName(name);
    } else {
      super.setName(id.substring(SCHEME.length()));
    }
  }
}
\end{lstlisting}

%
%


\newpage
\section{Finding 4: Top 100 Applications}
Regarding Finding 4, 
we have also looked into whether there was any relation between the number of contracts in the last version and any of the GitHub metrics from \autoref{fig:4-github-metrics}. However, no meaningful correlation was found.
\begin{figure}[h]
\centering
\begin{subfigure}{.5\columnwidth}
    \centering
    \includegraphics[width=1\columnwidth]{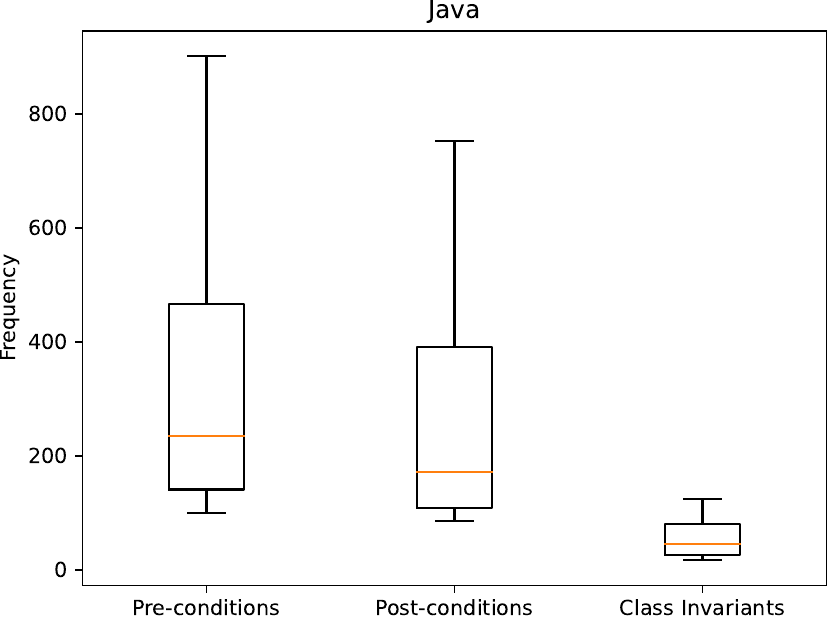}
    \label{fig:contractsByType_java}
\end{subfigure}%
\begin{subfigure}{.5\columnwidth}
    \centering
    \includegraphics[width=1\columnwidth]{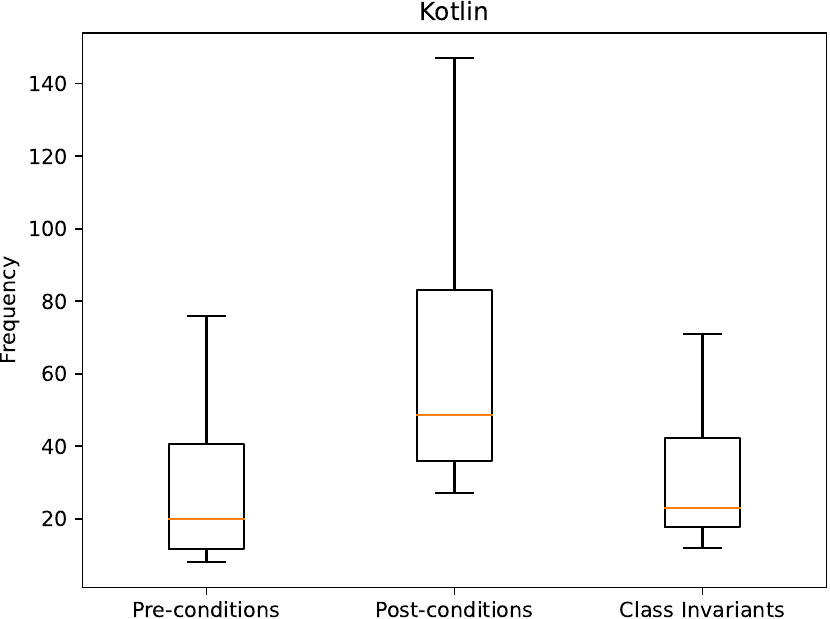}
    \label{fig:contractsByType_kotlin}
\end{subfigure}
\caption{Comparison of the distribution of the identified contract types in the top 100 applications with higher usage per type for Java and Kotlin.}
\label{fig:contractsByType_boxPlot}
\end{figure}

\newpage
\section{Table 6 extended}
In this section, we present a version of Table 6 (in the submitted paper) that also includes the number of applications (last two columns).

\begin{table*}[h]
\footnotesize
\renewcommand\arraystretch{1.1}
\centering
\caption{Number of contracts found in the dataset by construct and category.}
\begin{tabular}[t]{cccccccc}
    \hline
    \hline
     & & \multicolumn{2}{c}{contracts (all ver.)} & \multicolumn{2}{c}{contracts (2nd ver.)} & \multicolumn{2}{c}{applications} \\ 
     \cmidrule(lr){3-4} \cmidrule(lr){5-6} \cmidrule(lr){7-8}
    Construct & Category & Java & Kotlin & Java & Kotlin & Java & Kotlin \\
    \midrule
    cond. runtime exc. & CRE & 25,565 & 3,232 & 14,887 & 2,071 & 779 & 285 \\ 
    unsupp. op. exc. & CRE & 511 & 142 & 308 & 116 & 97 & 27 \\ 
    java assert & assertion & 3,525 & - & 2,217 & - & 325 & - \\ 
    kotlin assert & assertion & - & 3,868 & - & 2,370 & - & 234 \\ 
    guava precond. & API & 1,798 & 10 & 1,121 & 9 & 22 & 4 \\ 
    commons validate & API & 148 & 0 & 3 & 0 & 1 & 0 \\ 
    spring assert & API & 1 & 0 & 1 & 0 & 1 & 0 \\ 
    JSR303, JSR349 & annotation & 0 & 0 & 0 & 0 & 0 & 0 \\ 
    JSR305 & annotation & 4,195 & 20 & 2,133 & 13 & 40 & 4\\ 
    findbugs & annotation & 0 & 0 & 0 & 0 & 0 & 0 \\ 
    jetbrains & annotation & 2,310 & 138 & 1,596 & 98 & 115 & 20 \\ 
    android & annotation & 12,003 & 5,704 & 7,013 & 3,414 & 910 & 464 \\ 
    androidx & annotation & 139,933 & 20,593 & 86,212 & 13,811 & 599 & 401 \\ 
    kotlin contracts & others & - & 1 & - & 1 & - & 1 \\
    \bottomrule
\end{tabular}
\label{tab:contractsbytype}
\end{table*}

\section{List of Conditional Runtime Exceptions analyzed}

\begin{center}
\begin{longtable}{cc}
\caption{List of exceptions analyzed in the CRE category.} 
\label{tab:cre-exceptions}\\
\hline
\hline
\multicolumn{2}{c}{CREs Constructs} \\
\midrule 
AndroidRuntimeException             & MissingResourceException         \\
ArithmeticException                 & NegativeArraySizeException       \\
ArrayStoreException                 & NoSuchElementException           \\
ArrayStoreException                 & NullPointerException             \\
BufferOverflowException             & ParcelFormatException            \\
BufferUnderflowException            & ParseException                   \\
ClassCastException                  & ProviderException                \\
CompletionException                 & ProviderNotFoundException        \\
ConcurrentModificationException     & RejectedExecutionException       \\
DOMException                        & SQLException                     \\
DateTimeException                   & SecurityException                \\
EmptyStackException                 & TypeNotPresentException          \\
EnumConstantNotPresentException     & UncheckedIOException             \\
FileSystemAlreadyExistsException    & UndeclaredThrowableException     \\
FileSystemNotFoundException         & UnsupportedOperationException    \\
IllegalArgumentException            & WrongMethodTypeException         \\
IllegalMonitorStateException        & AcceptPendingException           \\
IllegalStateException               & AccessControlException          \\
IncompleteAnnotationException       & AlreadyBoundException            \\
IndexOutOfBoundsException             & AlreadyConnectedException        \\
LSException                         & ArrayIndexOutOfBoundsException    \\
MalformedParameterizedTypeException & BadParceableException            \\
MalformedParametersException        & CancellationException            \\
UnsupportedAddressTypeException     & UnsupportedCharsetException      \\
WritePendingException               & ZoneRulesException               \\
CancelledKeyException               & PatternSyntaxException           \\
ClosedDirectoryStreamException      & StringIndexOutOfBoundsException  \\
ClosedFileSystemException           & ReadOnlyBufferException          \\
ClosedFileSystemException           & ReadOnlyFileSystemException      \\
ClosedSelectorException             & ReadPendingException             \\
ClosedWatchServiceException         & ShutdownChannelGroupException    \\
ConnectionPendingException          & StringIndexOutOfBoundsException  \\
NonReadableChannelException         & UnknownFormatConversionException \\
NonWritableChannelException         & UnknownFormatFlagsException      \\
NotYetBoundException                & UnresolvedAddressException       \\
NotYetConnectedException            & UnsupportedTemporalTypeException \\
NumberFormatException               & OverlappingFileLockException  \\
\bottomrule
\end{longtable}
\end{center}
\section{List of API's methods analyzed}

\begin{center}
\begin{longtable}{cc}
\caption{List of the methods analyzed from each API.} 
\label{tab:api-methods-per-library}\\
\hline
\hline
\multicolumn{2}{c}{Annotations analyzed} \\
\midrule 
\multirow{7}{*}{Apache lang2 Validate} &\\
                           & allElementsOfType() \\
                           & isTrue() \\
                           & noNullElements() \\
                           & notEmpty() \\
                           & notNull() \\
                           &\\
\midrule
\multirow{11}{*}{Apache lang3 Validate} &  \\
                           & allElementsOfType() \\
                           & exclusiveBetween() \\
                           & inclusiveBetween() \\
                           & assignableFrom() \\
                           & isInstanceOf() \\
                           & matchesPattern() \\
                           & notBlank() \\
                           & validIndex() \\
                           & validState() \\
                           & \\
\midrule
\multirow[t]{8}{*}{Guava Preconditions}  & \\
                           & checkArgument() \\
                           & checkState() \\
                           & checkElementIndex() \\ 
                           & checkPositionIndex() \\
                           & checkNotNull() \\
                           & checkPositionIndexes() \\
                           & \\
\midrule
\multirow{13}{*}{Spring Assert}  & \\
                           & doesNotContain() \\
                           & hasLength() \\
                           & hasText() \\ 
                           & notEmpty() \\
                           & noNullElements() \\
                           & isInstanceOf() \\
                           & isAssignable() \\
                           & state() \\
                           & isNull() \\
                           & isTrue() \\
                           & notNull() \\
                           & \\
\bottomrule
\end{longtable}
\end{center}
\section{List of Annotations analyzed}

\begin{center}
\begin{longtable}{m{2cm}cc}
\caption{List of annotations analyzed per package.} 
\label{tab:annotations-per-packages}\\
\hline
\hline
\multicolumn{3}{c}{Annotations analyzed} \\
\midrule 
\multirow{13}{*}{JSR305}   & & \\
                           & @CheckForNull & @CheckForSigned \\
                           & @MatchesPattern & @Nonnegative \\
                           & @Nonnull & @Nullable\\
                           & @OverridingMethods\-MustInvokeSupper & @ParametersAre\-NonnullByDefault\\
                           & @RegEx & @Signed\\
                           & @Syntax & @Syntax \\
                           & @Tainted & @Untainted\\
                           & @WillClose & @WillCloseWhenClosed\\
                           & @WillNotClose & @Guardedby \\
                           & @Immutable & @NotThreadSafe \\
                           & @ThreadSafe & \\
                           & & \\
\midrule
\multirow{9}{*}{JSR303, JSR349} & & \\   
                           & @Null & @DecimalMin \\
                           & @NotNull & @Size \\
                           & @AssertTrue & @Digits\\
                           & @AssertFalse & @Past\\
                           & @Min & @Future\\
                           & @Max & @Pattern \\
                           & @DecimalMax & \\
                           & & \\
\midrule
\multirow[t]{5}{*}{JetBrain}  & & \\
                           & @Contract & @NotNull \\
                           & @Nullable & @PropertyKey \\
                           & @TestOnly & \\
                           & & \\
\midrule
\multirow{14}{*}{IntelliJ} & & \\
                           & @BoxLayoutAxis & @CalendarMonth\\
                           & @CursorType & @FlowLayoutAlignment \\
                           & @FontStyle & @HorizontalAlignment\\
                           & @InputEventMask & @ListSelectionMode\\
                           & @PatternFlags & @TabLayoutPolicy\\
                           & @AdjustableOrientation & @Flow\\
                           & @Identifier & @TabPlacement\\
                           & @TitledBorderJustification & @TitledBorderTitlePosition\\
                           & @Language & @MagicConstant\\
                           & @Pattern & @PrintFormat\\
                           & @PrintFormat & @RexExp\\
                           & @Subst & \\
                           & & \\
\midrule
\multirow{7}{*}{FindBugs} & & \\
                           & @CheckForNull & @NonNull\\
                           & @Nullable & @PossiblyNull \\
                           & @FontStyle & @HorizontalAlignment\\
                           & @UnkownNullness & @CreateObligation\\
                           & @DischargesObligation & @CleanupObligation\\
                           & & \\
\midrule
\multirow{3}{*}{Android} & & \\
                           & @AndroidSupressLint & @AndroidTargetApi\\
                           & & \\
\midrule
\multirow[t]{32}{*}{Androidx} & & \\
                           & @AnimatorRes & @AnimRes\\
                           & @AnyRes & @AnyThread \\
                           & @AnyThread & @ArrayRes \\
                           & @AttrRes & @BinderThread \\
                           & @BinderThread & @BoolRes \\
                           & @CallSuper & @CheckResult \\
                           & @ChecksSdkIntAtLeast & @ColorInt \\
                           & @ColorLong & @ColorRes \\
                           & @ContentView & @DimenRes \\
                           & @Dimension & @NotInline \\
                           & @DrawableRes & @FloatRange \\
                           & @FloatRange & @FontRes \\
                           & @FontRes & @FractionRes \\
                           & @FractionRes & @GuardedBy \\
                           & @GuardedBy & @HalfFloat \\
                           & @IdRes & @InspectableProperty \\
                           & @IntDef & @IntegerRes \\
                           & @InterpolatorRes & @IntRange \\
                           & @Keep & @LayoutRes \\
                           & @LongDef & @MainThread \\
                           & @MainThread & @MenuRes \\
                           & @NavigationRes & @NonNull \\
                           & @Nullable & @PluralsRec \\
                           & @Px & @RawRes \\
                           & @RequiresApi & @RequiresFeature \\
                           & @RequiresPermission & @RestrictTo \\
                           & @Size & @StringDef \\ 
                           & @StringRes & @StyleableRes \\
                           & @StyleRes & @TransitionRes \\
                           & @UiThread & @VisibleForTesting \\
                           & @WorkerThread & @XmlRes \\
\bottomrule
\end{longtable}
\end{center}

\end{document}